\documentclass[useAMS,usenatbib]{mn2e}
\usepackage{amssymb}
\usepackage{graphicx}
\usepackage{url}
\usepackage[usenames]{color}
\usepackage{url}
\usepackage[percent]{overpic}

\newcommand{\de} {{\rm d}}
 
\begin{document}

\title[$\gamma$-ray spectra of Geminga, Crab, and Vela]{Modelling of the $\gamma$-ray pulsed spectra of Geminga, Crab, and Vela with synchro-curvature radiation}

\author[D.~Vigan\`o \& D.~F.~Torres]{Daniele Vigan\`o$^1$ \& Diego F.~Torres$^{1,2}$\\ 
$^1$Institute of Space Sciences (CSIC--IEEC), Campus UAB, C. de Can Magrans, s/n 08193, Cerdanyola del Valles (Barcelona), Spain\\
$^2$Instituci\'o Catalana de Recerca i Estudis Avan\c{c}ats (ICREA), 08010, Barcelona, Spain}

\date{}
\maketitle

\label{firstpage}

\begin{abstract}
$\gamma$-ray spectra of pulsars have been mostly studied in a phenomenological way, by fitting them to a cut-off power-law function. Here, we analyse a model where pulsed emission comes from synchro-curvature processes in a gap. We calculate the variation of kinetic energy of magnetospheric particles along the gap and the associated radiated spectra, considering an effective particle distribution. We fit the phase-averaged and phase-resolved {\em Fermi}-LAT (Large Area Telescope) spectra of the three brightest $\gamma$-ray pulsars: Geminga, Crab, and Vela, and constrain the three free parameters we leave free in the model. Our best-fitting models well reproduce the observed data, apart from residuals above a few GeV in some cases, range for which the inverse Compton scattering likely becomes the dominant mechanism. In any case, the flat slope at low-energy ($\lesssim$ GeV) seen by {\it Fermi}-LAT both in the phase-averaged and phase-resolved spectra of most pulsars, including the ones we studied, requires that most of the detected radiation below $\sim$GeV is produced during the beginning of the particle trajectories, when radiation mostly come from the loss of perpendicular momentum. 
\end{abstract}

\section{Introduction}

The wealth of {\em Fermi}-LAT (Large Area Telescope) data \citep{2fpc} has boosted our knowledge about $\gamma$-ray emission from pulsars, allowing a better understanding of the fundamental high-energy processes which are responsible for the conversion of the rotational energy into radiation. Two main channels are the likely origin of the detected radiation: photons emitted by particles moving in curved magnetic fields and accelerating electric fields, i.e., the synchro-curvature (SC) radiation, and the inverse Compton (IC) scattering of background photons against energetic magnetospheric particles \citep{bogovalov00}.

This paper is the continuation of a series of works \citep{paper0,paper1,paper2}, in which we focus on the high-energy SC radiation, and to which we refer the reader for further details. Our approach applies to any general gap located in the outer magnetosphere (even outside the light cylinder), since our parameters do not depend on an a-priori, detailed choice of the location of the gap. Instead, we show here how data can be used to constrain the values of the most relevant gap parameters of the model. 

The $E^2dN/dE$ spectra of pulsed $\gamma$-ray emission from most pulsars peak around a few GeV, above which the flux quickly decreases with energy. Spectra are usually described by a cut-off power-law:
\begin{equation}\label{eq:cutoff}
  \frac{dP}{dE} = P_0 E^\mu \exp\left[-\left(\frac{E}{E_p}\right)^s\right]~,
\end{equation}
where the four parameters to be fitted are the normalization $P_0$, the low-energy slope index $\mu$, the peak energy $E_p$, and the exponential index $s$. The values of the best-fitting models give a $\mu$-range in the interval $(-1, 0.5)$, with the distribution peaking close to $\mu \sim -0.5$  (see Fig.~7 of \citealt{2fpc}, where their $\Gamma$ is related to $\mu$ as $\Gamma = 1 - \mu$). This relative flatness at low energies (0.1-1 GeV) contrasts with the value of $\mu$ predicted by the SC spectrum of a single-particle, which has a value $\mu = 0.25$ due to mathematical properties of the functions involved. The latter is consistent only with a small minority of observed pulsars. This fact means that a simple SC radiation model which considers a mono-energetic distribution of particles is unable on a first-principle-basis to explain most of the observed spectra.

The relative flatness (i.e., $\mu \sim -1$) of many pulsar spectra at low {\it Fermi}-LAT energies is a basic issue that to our knowledge has never been properly addressed. In particular, we believe it could signal that non-saturated particles are important contributors to the total emitted radiation. Put otherwise, that the contribution of different parts of the particle trajectories is non-uniform. In \cite{paper0,paper2} we showed how a large weight given to the initial parts of the trajectories, where the radiation is dominated by synchrotron-like emission, can explain flatter slopes ($\mu < 0.25$). We study this possibility in more detail here, and advance that it is in agreement with data.

Pulsar spectra show another important feature: for some sources, the phase-averaged spectrum shows a sub-exponential cut-off, i.e., $s < 1$. The high-energy tail ($E\gtrsim$ few GeV) decreases in energy slower than the expected SC radiation emitted by mono-energetic particles, which produces a purely exponential cut-off, $s=1$. This issue alone does not rule out SC radiation as the dominant mechanism: the phase-averaged spectrum is the superposition of radiation coming from different phases, each of which could be properly described by a SC spectrum. The superposition of such spectra, with different energy peaks, slopes and fluxes can well lead to a sub-exponential cutoff. Such scenario is found for the Geminga pulsar (see \citealt{abdo10c} and the discussion below). The study of the phase-resolved spectra is the only way to assess this issue, and possibly rule out the SC radiation as the dominant mechanism above a few GeV. 

In this work, we calculate the variation of energy and the radiated spectra of magnetospheric particles along a gap, considering an effective particle distribution. We fit the phase-averaged and phase-resolved {\em Fermi}-LAT spectra of the three brightest $\gamma$-ray pulsars: Geminga, Crab, and Vela. With such models we address the above mentioned issues. In \S\ref{sec:spectrum} we recall the formulae describing the SC radiation, the computation of the particle trajectories and the associated radiative losses spectra. In \S\ref{sec:results} we fit the phase-averaged and phase-resolved spectra of the radio-quiet PSR~J0633+1746 (hereafter, Geminga as it is also known; \citealt{halpern92}), PSR~J0534+2200 (the Crab pulsar; \citealt{staelin68}), and PSR~J0835--4510 (the Vela pulsar; \citealt{large68}). By comparing models with data, we constrain the key physical parameters of our gap models, in particular, the component of the electric field parallel to the magnetic field (hereafter, $E_\parallel$) and the effective particle distribution. In \S\ref{sec:conclusions}, we discuss the obtained results and its implications.

%In a future work, we plan to systematically study the good-quality phase-averaged spectra of the pulsars, in a population study approach.

\section{SC radiation}\label{sec:spectrum}

\subsection{Single-particle radiation}

The SC radiation emitted by a single, charged particle spiralling around a magnetic field line is described by the following formula for the energy spectrum (see \citealt{cheng96} and the subsequent reformulation by \citealt{paper0}):
\begin{equation}\label{eq:sed_synchrocurv}
 \frac{dP_{\rm sc}}{dE} = \frac{\sqrt{3} e^2 \Gamma y}{4\pi \hbar r_{\rm eff} } [ (1 + z) F(y) - (1 - z) K_{2/3}(y)]~,
\end{equation}
where
\begin{eqnarray}
 && z= (Q_2 r_{\rm eff})^{-2} ~, \label{eq:z}\\
 && F(y) = \int_y^\infty K_{5/3}(y') dy'~,\label{eq:f_y}\\
 && y=\frac{E}{E_c} ~, \\
 && E_c = \frac{3}{2}\hbar cQ_2\Gamma^3~,\label{eq:echar}\\
 && r_{\rm gyr} = \frac{mc^2\Gamma\sin\alpha}{eB}~,\\
 && Q_2 = \frac{\cos^2\alpha}{r_c}\sqrt{1 + 3\xi  + \xi^2 + \frac{r_{\rm gyr}}{r_c}} \label{eq:q2}~, \\
 && \xi = \frac{r_c}{r_{\rm gyr}}\frac{\sin^2\alpha}{\cos^2\alpha}~,\\
 && r_{\rm eff} = \frac{r_c}{\cos^2\alpha}\left(1 + \xi+ \frac{r_{\rm gyr}}{r_c}  \right)^{-1}~,\\
 && g_r =  \frac{r_c^2}{r_{\rm eff}^2}\frac{[1 + 7(r_{\rm eff}Q_2)^{-2}]}{8 (Q_2r_{\rm eff})^{-1}}~.\label{eq:gr}
\end{eqnarray}
In the formulae above, $m$ and $\Gamma$ are the rest mass and the Lorentz factor of the particle, $\alpha$, $r_{\rm gyr}$ and $r_c$ are the pitch angle, the Larmor radius, and the radius of curvature of its trajectory, respectively, $e$ the elementary charge, $B$ the local strength of the magnetic field, $\hbar$ the reduced Planck constant, $c$ is the speed of light, $K_n$ are the modified Bessel functions of the second kind of index $n$, $E$ is the photon energy, $E_c$ is the characteristic energy of the emitted radiation. The factors $r_{\rm eff}$, $Q_2$, $\xi$, and $g_r$ are introduced to provide a compact formula. In the limits of high or vanishing perpendicular momentum, Eq.~(\ref{eq:sed_synchrocurv}) reduces to purely synchrotron ($\xi\gg1$) or curvature radiation ($\xi\ll 1$) formulae, respectively. In any case, the peak of the spectrum is located close to $E_c$ and, for energies $E\ll E_c$, the dominant term in Eq.~(\ref{eq:sed_synchrocurv}), $F(y)$, provides $dP_{\rm sc}/dE \sim E^{0.25}$. See \cite{paper0} for a more detailed discussion.

\subsection{SC radiation from the pulsar magnetosphere}

The detectable radiation coming from the magnetosphere can be estimated by integrating the single-particle spectrum, Eq.~(\ref{eq:sed_synchrocurv}), along the travelled distance between the assumed boundaries $x_{\rm in}$ and $x_{\rm out}$, convoluted with an effective particle distribution, $dN/dx$, where $x$ is the distance along the magnetic field line \citep{paper2}:
\begin{equation}\label{eq:sed_x}
  \frac{dP_{\rm gap}}{dE_\gamma} =  \int_{x_{\rm in}}^{x_{\rm out}} \frac{dP_{\rm sc}}{dE_\gamma}\frac{dN}{d x} {\rm d}x~.
\end{equation}
The effective particle distribution represents the number of particles, per unit of distance, emitting radiation towards us. In  general, it differs from the total particle distribution which is very hard to be inferred on purely theoretical basis.

There exist two possible approaches to infer such a distribution. The first one is a semi-analytical one, followed by e.g., \cite{zhang97}, in which most gap parameters are univocally determined by the values of $P$ and $B_\star$. In \cite{paper1}, we questioned such approach, analysing the impact of its many underlying assumption and approximations, which necessarily relies on, for instance, taking single-value parameters in many quantities, neglecting their uncertainties and dispersions (caused, e.g., by viewing angle, beaming effects and the magnetospheric geometry), or fixing the pitch angle to a non-zero value \citep{zhang97}, while it should go to zero under SC losses \citep{paper0}.

On the other hand, the numerical simulations by \cite{takata06} and \cite{hirotani06,hirotani07,hirotani08} include the pair cascade, the Boltzmann equation for particles and the interaction with the radiation. These simulations are physically more self-consistent, but they also rely on some intrinsic, hardly avoidable assumptions, like the precise location of the gap, the intensity of incoming/outgoing currents (if any), and a simplified magnetic field geometry. Moreover, the computational cost makes this approach unviable if one wants to fit real data once and again. More recent state-of-the-art MHD simulations \citep{kalapotharakos14,brambilla15} roughly predict the curvature radiation emitted outside the light cylinder, under certain assumptions (negligible pitch angle, spin period of $\sim$ ms, constant magnetospheric conductivity outside the light cylinder, and purely force-free magnetosphere inside it).

\begin{figure}
\centering
\includegraphics[width=.42\textwidth]{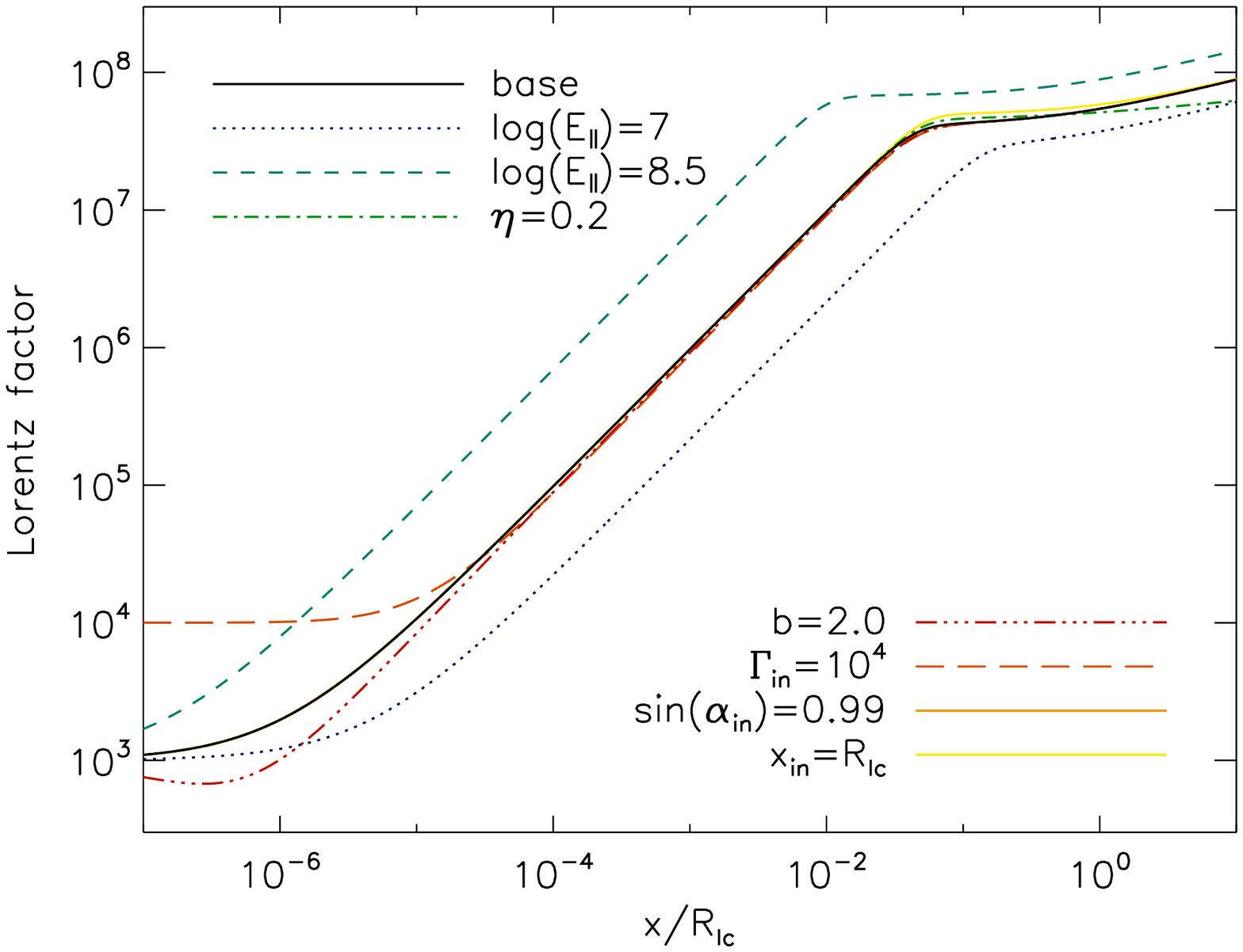}\\
\includegraphics[width=.42\textwidth]{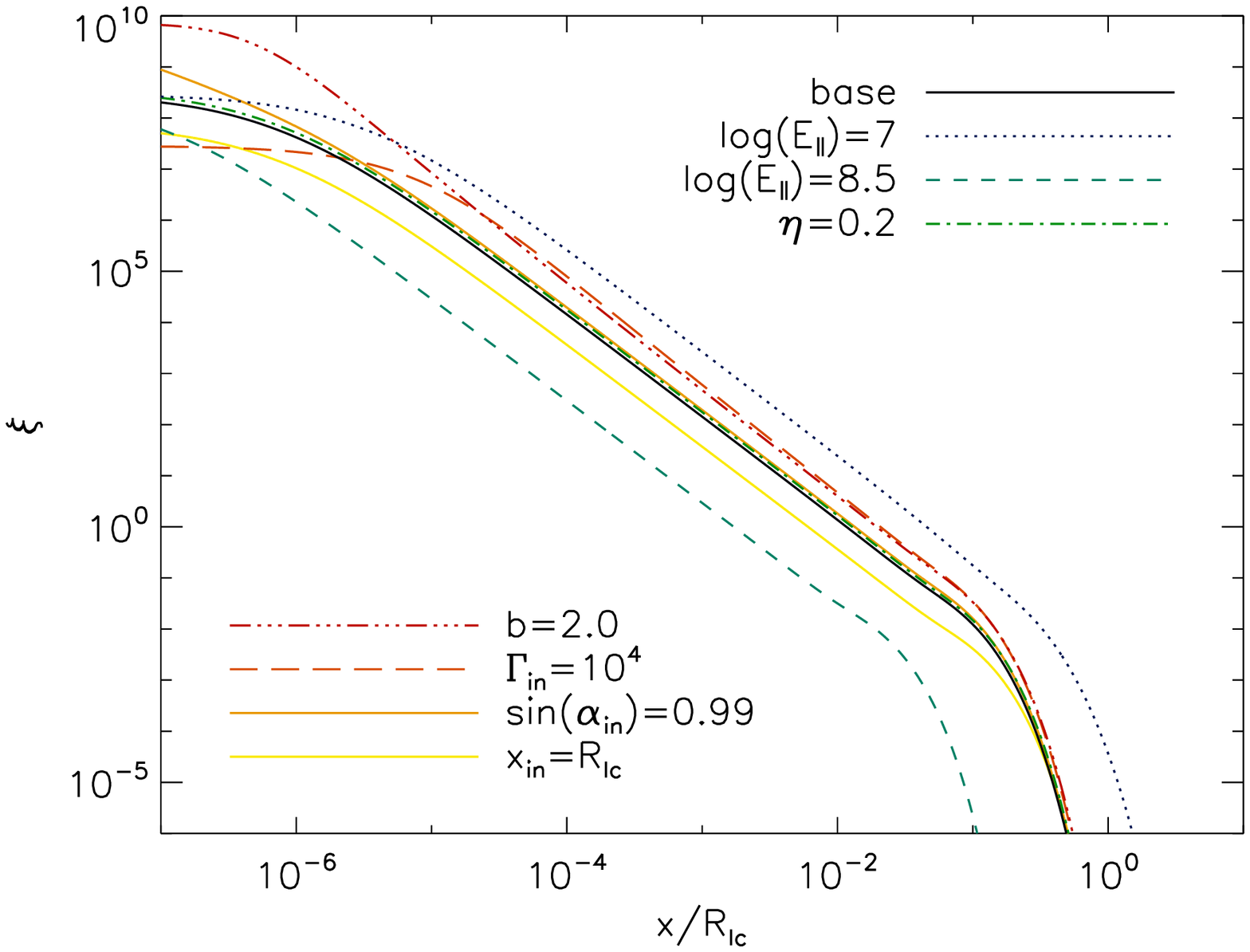}\\
\includegraphics[width=.42\textwidth]{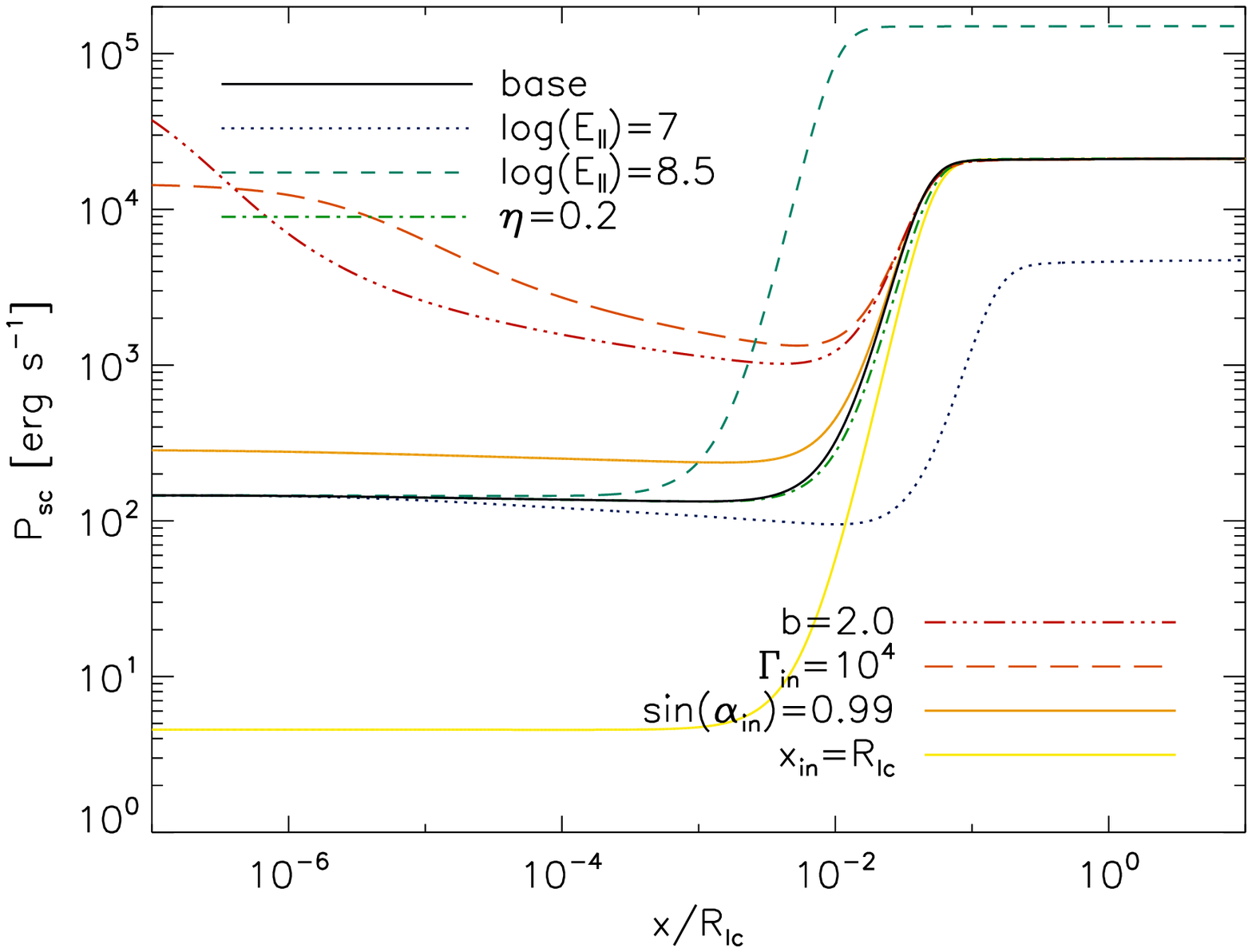}
\caption{Evolution of the Lorentz factor $\Gamma$ (top), the SC parameter $\xi$ (middle), and the single-particle SC power $P_{\rm sc}$ (bottom) calculated for a model with the timing properties of Geminga pulsar (see Table~\ref{tab:pulsars}), and varying one by one the relevant parameters, as indicated in the legend, over a baseline model consisting of: $E_\parallel=10^{7.60}$ V~$m^{-1}$, $\eta=0.5$, $b=2.5$, $\Gamma_{\rm in}=10^3$, $\alpha_{\rm in}=\pi/4$, $x_{\rm in}/R_{\rm lc}=0.5$. Some of the lines overlap with the base model (black solid line).}
 \label{fig:trajectories}
\end{figure}

\begin{table}
\begin{center}
\caption{Parameters entering in the SC model, explored range, and treatment in the fitting procedure.  For a given pulsar, $P$ and $B_\star$ are fixed by timing observed properties. The second group of three parameters are the ones constrained by data (we indicate within parenthesis which is the most constraining spectral property), while the remaining six parameters have much smaller influence on spectra and are fixed to the listed values.}
\label{tab:parameters}
\begin{tabular}{c c l}
\hline
\hline
Parameter & Range & Treatment\\
\hline
$P$ & - & timing data \\
$B_\star$ & - & fixed by timing data \\
\hline
$\log(E_\parallel[{\rm V~m^{-1}}])$ & 6.5-9.5 & fit (energy peak) \\
$N_0$ & $10^{26}$--$10^{34}$ & fit (luminosity) \\
$x_0/R_{\rm lc}$ & 0.001-1$^a$ & fit (low-energy slope) \\
\hline
$\eta$ & 0.2-1.0 & fixed to 0.5 \\
$b$ & 2-3 & fixed to 2.5 \\
$x_{\rm in}/R_{\rm lc}$ & 0.2-1.0 & fixed to 0.5 \\
$x_{\rm out}/R_{\rm lc}$ & 1.0-2.0 & fixed to 1.5 \\
$\Gamma_{\rm in}$ & $10^3$-$10^4$ & fixed to $10^3$ \\
$\alpha_{\rm in}$ & $0$-$\pi/2$ & fixed to $\pi/4$ \\
\hline
\hline
\end{tabular}
\end{center}
$^a$ See Eq.~(\ref{eq:distribution}) for the definition of the effective parameter distribution $dN/dx$. We also explore the uniform particle distribution, $dN/dx = N_0/(x_{\rm out}-x_{\rm in})$.
\end{table} 

These difficulties in fitting spectra from first principles led us to take a more effective approach, based on a gap model containing a few effective parameters, which allow a systematic study of $\gamma$-ray spectra from which to extract few, but solid, physical conclusions. We explore here the effects of having effective particles distributions with larger weights in the inner part of the trajectory (small $x_0/R_{\rm lc}$, see below), as discussed in \cite{paper0,paper2}. This choice reflects both an observational and a theoretical consideration. As said in the introduction (see also \citealt{paper2}) a uniform distribution is not compatible with having the variety of slopes at low energies, $E\lesssim $ GeV, seen by {\it Fermi}-LAT. Physically, we can consider a number of effects that can qualitatively justify a large  weight in the low $\xi$ parts of the trajectories: the beaming is larger at smaller $\Gamma$, so that it is easier to detect radiation for less energetic particles. The geometry of the emitting region can also affect the radiation. The cascade of pairs, produced especially in the inner gap, where the interactions with X-ray photons are more likely, can also give a net effect of enhancing visible emission from particles with $\xi \gg 1$, when they lose perpendicular momentum. Other effects, related to magnetic geometry and viewing angle, can further complicate the picture, and make any theoretical prediction quite difficult.

In order to introduce such effective particle distribution in a way that allow fitting, we shall introduce a functional form depending on the length-scale $x_0$ and normalization $N_0$ (total effective number of particles):

\begin{equation}  \label{eq:distribution}
  \frac{dN}{dx}=N_0\frac{e^{-(x-x_{\rm in})/x_0}}{x_0(1 - e^{-x_{\rm out}/x_0}) }~.
\end{equation}
The choice of this particular function is motivated by simplicity. Other functions, with other parameters allowing to produce a non-uniform particle distribution along the gap could have been considered. We stress that this choice is an effective way to parametrize our ignorance about the combined effects described above on the observed emission, and will allow to constrain significant qualitative features emerging from the data fits. We also stress that Eq.~(\ref{eq:distribution}) shall represent the particles which effectively produce the observed radiation, not the whole particle population in the gap, which we may never infer.

\subsection{Particles trajectories}

\begin{table*}
\begin{center}
\caption{Pulsars considered in this work: timing properties, references for the spectral analysis, phases considered (average, peaks and minimum or inter-pulse state), and the values of the relevant parameters of our SC best-fiting models, with the associated value of $\tilde{\chi}^2$. The loss of the rotational energy is calculated as $\dot{E}_{\rm rot}= 4\pi I \dot{P}/P^3=3.9\times 10^{46} \dot{P}/P{\rm [s]}^3$ erg/s, while the characteristic age is defined as $\tau=P/2\dot{P}$.}
\label{tab:pulsars}
\begin{tabular}{c c c c c c c c c c c c}
\hline
\hline
Pulsar & $P$ & $\dot{P}$ & $B_\star$ & $\tau$ & $\dot{E}_{\rm rot}$ & Refs. & Phases & $\log(E_\parallel)$ & $x_0$ & $N_0$ & $\tilde{\chi}^2$ \\
 & [s] & $ [10^{-14}] $ & [$10^{12}$G] & [yr] & [erg/s] & & &  [V $m^{-1}$] & [$R_{\rm lc}$] & [$10^{30}$] & \\
\hline
J0633+1746 & 0.237 & 1.1 & 3.3 & $1.2\times 10^3$ & $3.2\times 10^{34}$ & [1] & average & 7.65 & 0.013 & 18.8 & 18.1 \\
(Geminga) & & & & & & & P1: 0.637-0.643 & 7.80 & 0.013 & 0.26 & 0.68  \\ % ph28
& & & & & & & P2: 0.131-0.141 & 7.65 & 0.010 & 0.82 & 0.91 \\ % ph05
& & & & & & & Min.: 0.942-1.000 & 7.20 & 0.026 & 0.76 & 1.17 \\ % ph35
\hline
J0534+2200 & 0.033 & 42.3 & 7.6 & $1.1\times 10^4$ & $4.6\times 10^{38}$ & [2] & average & 8.75 & 0.003 & 41.3 & 4.55 \\ 
(Crab)& & & & & & & P1: 0.987-0.993 & 8.60 & 0.004 & 3.16 & 1.48 \\ % ph13
& & & & & & & P2: 0.366-0.386 & 8.80 & 0.003 & 1.8 & 1.48 \\ % ph06
& & & & & & & Bridge: 0.098-0.286 & 8.45 & 0.010 & 2.9 & 0.54 \\ % ph03
\hline
J0835--4510 & 0.089 & 12.5 & 6.8 & $3.4\times 10^5$ & $6.9\times 10^{36}$ & [3] & average & 8.05 & 0.010 & 14.6 & 136 \\
(Vela) & & & & & & & P1: 0.133-0.135 & 8.05 & 0.008 & 0.22 & 2.12 \\ % ph01
& & & & & & & P2: 0.562-0.563 & 8.00 & 0.014 & 0.13 & 2.76 \\ % ph04
& & & & & & & P3: 0.315-0.324 & 8.15 & 0.013 & 0.076 & 1.39 \\ % ph03
\hline
\hline
\end{tabular}
\end{center}
\begin{minipage}{\textwidth}
References: [1] \cite{abdo10c}; [2] \cite{abdo10a}; [3] \cite{abdo10b}.
\end{minipage}
\end{table*}

The quantities defined in Eqs.~(\ref{eq:z})-(\ref{eq:gr}), appearing in Eq.~(\ref{eq:sed_synchrocurv}), depend on the local values of $\Gamma$, $\alpha$, $r_c$, and $B$. The kinematic parameters, $\Gamma$ and $\sin\alpha$, evolve along the gap due to electric acceleration and radiation losses, thus we simulate the motion of charged particles by evolving the parallel and perpendicular momenta of particles according to the equations of motion \citep{paper0}:
\begin{eqnarray}
 && \frac{\de(p\sin\alpha)}{\de t} = - \frac{P_{\rm sc}\sin\alpha}{v}~, \label{eq:motion_perp} \\
 && \frac{\de(p\cos\alpha)}{\de t} = eE_\parallel - \frac{P_{\rm sc}\cos\alpha}{v}~, \label{eq:motion_par}
\end{eqnarray}
where $p=\Gamma m v$ is the momentum of the particle, $v$ its spatial velocity, $E_\parallel$ is the parallel component of the electric field, considered constant along the gap \citep{paper1}, and $P_{\rm sc}$ is the single-particle SC power, obtained by integrating Eq.~(\ref{eq:sed_synchrocurv}) in energy:
\begin{equation}\label{eq:power_synchrocurv}
 P_{\rm sc} = \frac{2e^2 \Gamma^4 c}{3 r_c^2} g_r~.
\end{equation}
On the other hand, we prescribe simple functional forms of $r_c(x)$ and $B(x)$, depending on two effective parameters, $\eta$ and $b$, as justified in \cite{paper1,paper2}:
\begin{eqnarray}
 && r_c(x) = R_{\rm lc}\left(\frac{x}{R_{\rm lc}}\right)^\eta~, \\
 && B(x) = B_\star \left(\frac{R_\star}{x}\right)^b~,
\end{eqnarray}
where $R_\star$ is the neutron star (NS) radius, and we employ the standard definitions of light cylinder distance and inferred dipolar magnetic field at the polar surface:
\begin{eqnarray}
 && R_{\rm lc} = \frac{Pc}{2\pi} \simeq 4.77\times 10^9 P{\rm[s] ~ cm} ~,\label{eq:rlc} \\
 && B_\star = 6.4 \times 10^{19} ~ \sqrt{P{\rm [s]}\dot{P}}~{\rm ~G}~. \label{eq:bstar}
\end{eqnarray}
Expected values for a pulsar magnetosphere are $\eta \sim 0.2-1$ and $b\sim 2-3$ \citep{paper1}. 

We summarize here the main results obtained in our previous works (\citealt{paper0,paper2}, see also \citealt{hirotani99a}) in Fig.~\ref{fig:trajectories}, by showing the evolution of the Lorentz factor $\Gamma$ (top), the SC parameter $\xi$ (middle), and the single-particle SC power $P_{\rm sc}$ (bottom), obtained for models having the timing parameters of the Geminga pulsar. Soon after pair creation, $\xi \gg 1$, i.e. the losses mostly regard the perpendicular momentum, being approximated by synchrotron formula. Then, in a length-scale $x\ll R_{\rm lc}$, the electrical acceleration makes $\Gamma$ increase and $\sin\alpha$ decrease by orders of magnitude, until the radiative losses can be approximated by purely curvature radiation, $\xi \ll 1$. In this regime, $\Gamma$ and $P_{\rm sc}$ tends to saturate to a value determined by the balance between the electric force and the radiation-reaction force.

The baseline model (black line) has the following parameters: $P=0.237$ s, $B_\star=3.3\times 10^{12}$ G (as in the Geminga pulsar), $E_\parallel=10^{7.6}$ V~m$^{-1}$, $\eta=0.5$, $b=2.5$, $x_{\rm in}=0.5 R_{\rm lc}$, $\Gamma_{\rm in}=10^3$, and $\alpha_{\rm in}=\pi/4$. The other lines show models obtained exploring reasonable variations of the relevant parameters, according to what studied in \cite{paper1}. $b$, $\Gamma_{\rm in}$, and $\alpha_{\rm in}$ have visible effects only in the very early part of the trajectories, $x/R_{\rm lc} \lesssim 10^{-5}$. This happens because, when $\xi \lesssim 1$, then the radiation is approximated by curvature radiation, which is independent on the value of $B$ and $\alpha$. The parameters $x_{\rm in}$ and $\eta$ have an overall negligible influence as well, limited to minor differences in the outer part of the trajectory. Hence, the most relevant parameter affecting the trajectories is, by far, $E_\parallel$: the larger it is, the larger $\Gamma$ and the smaller $\xi$.

In conclusion, we can see that for $x/ R_{\rm lc} \lesssim 10^{-3}-10^{-1}$ (value depending on the model) the deviations from a purely curvature radiation ($\xi \ll 1$) can be important. This is why we need to consider the full expression of SC radiation and evaluate it along the whole trajectory of the particles moving through the gap, especially if most of the detectable radiation originates at the beginning of their trajectories, when particles are relatively slow and radiate mostly due to perpendicular momentum losses.

\begin{figure*}
\begin{overpic}[width=0.36\textwidth]{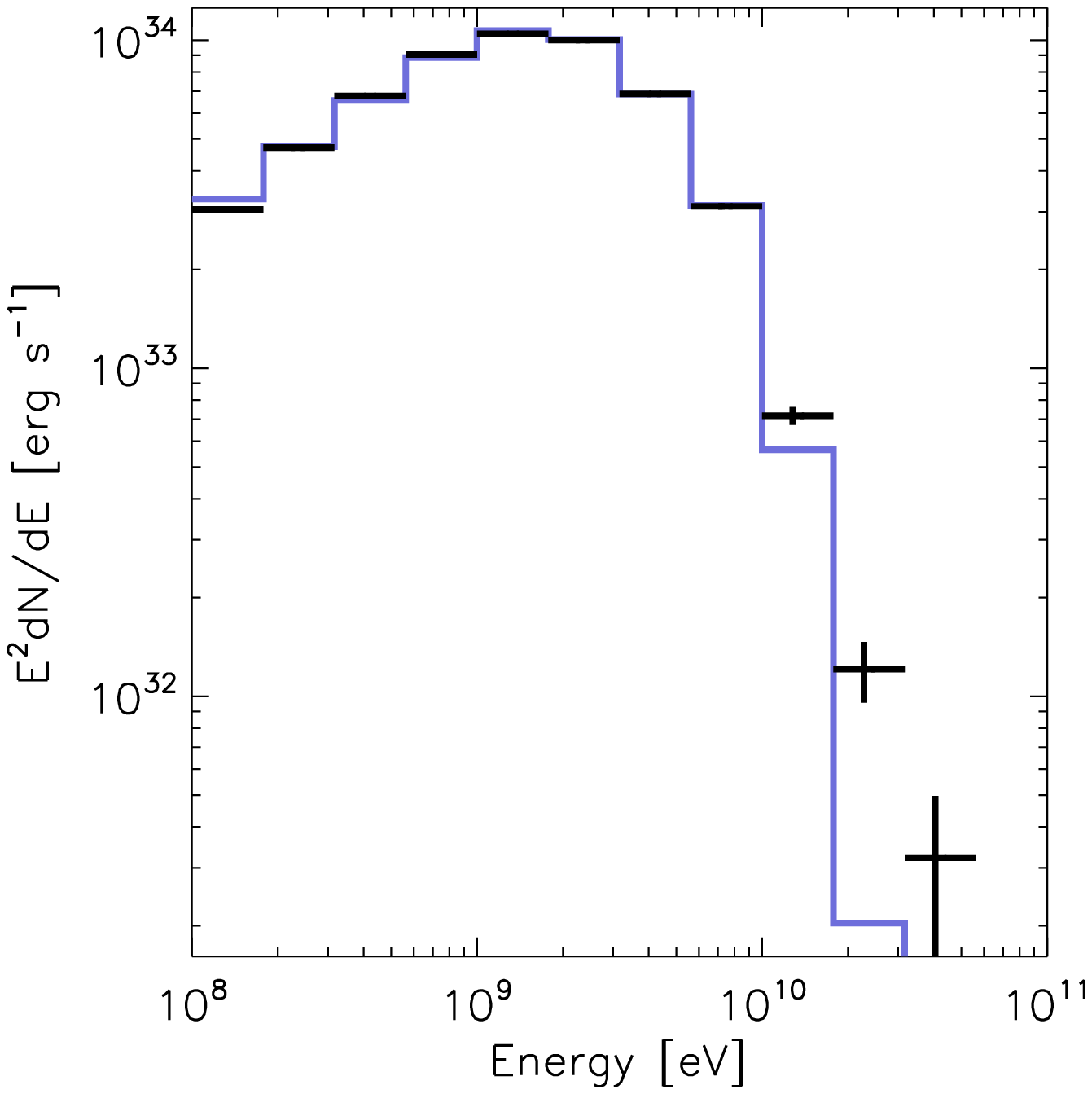}
\put(45,95){Geminga average}
\end{overpic}
\begin{overpic}[width=0.36\textwidth]{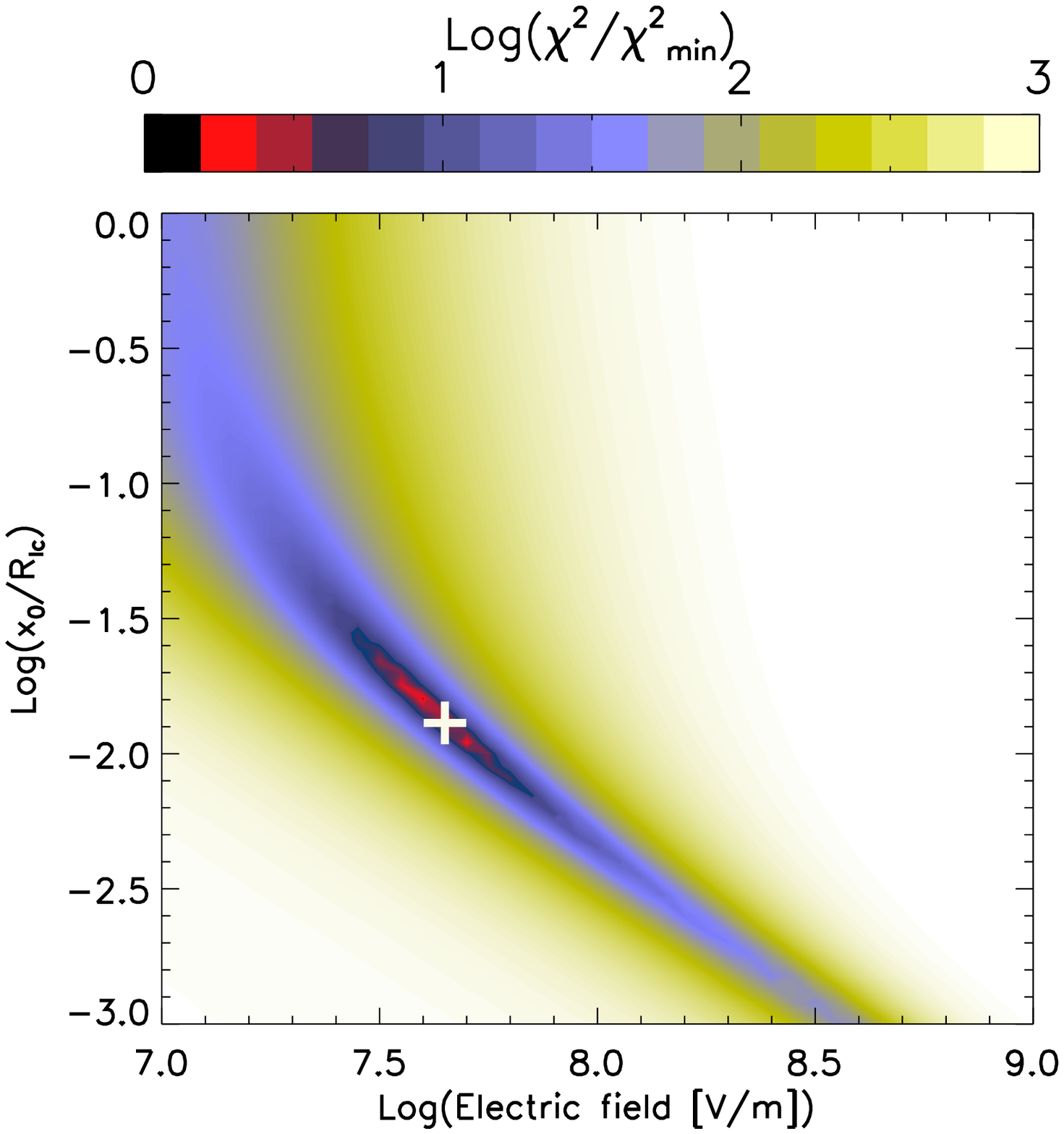}
\put(60,75){Geminga average}
\end{overpic}
\includegraphics[width=.32\textwidth]{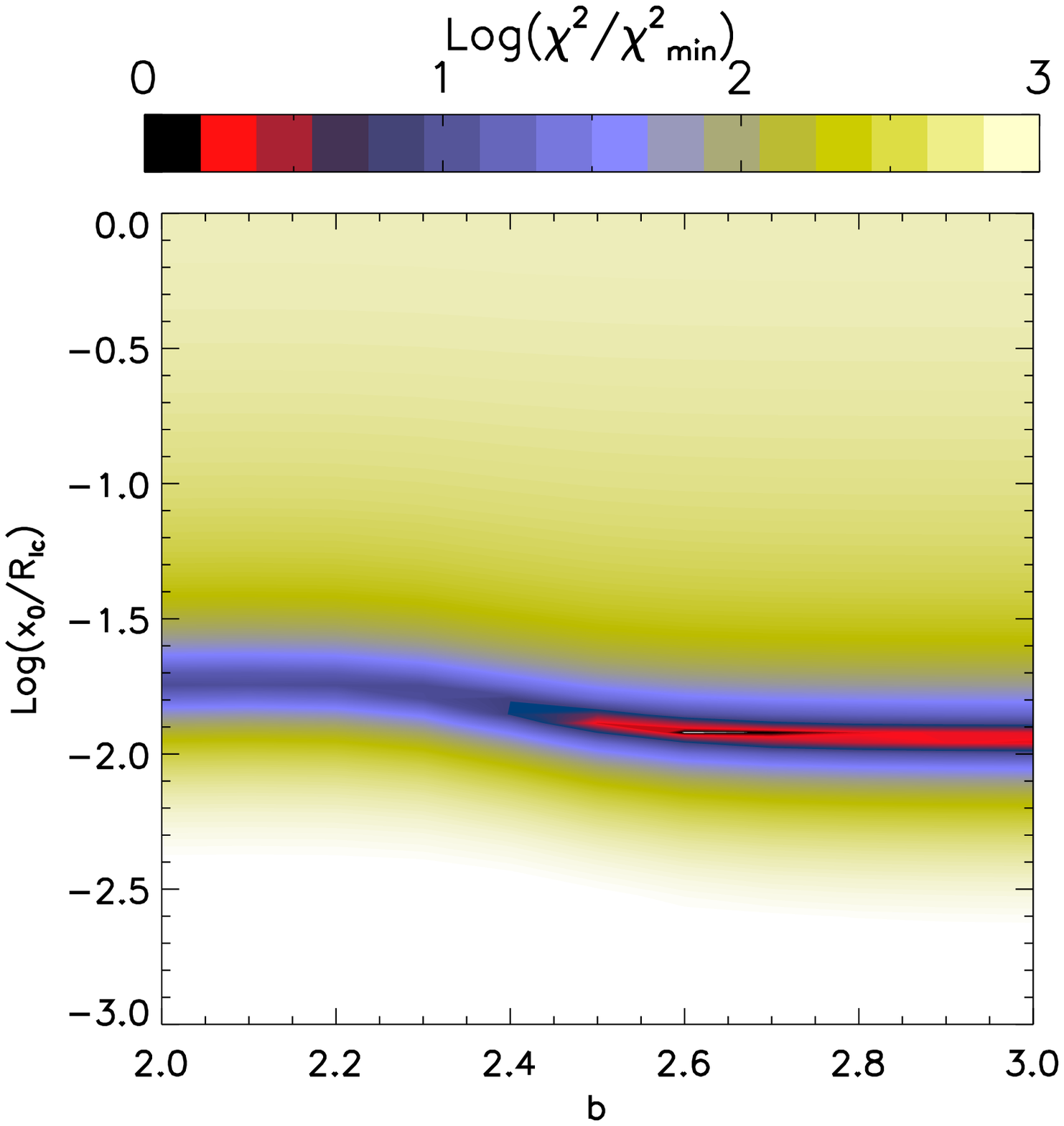}
\includegraphics[width=.32\textwidth]{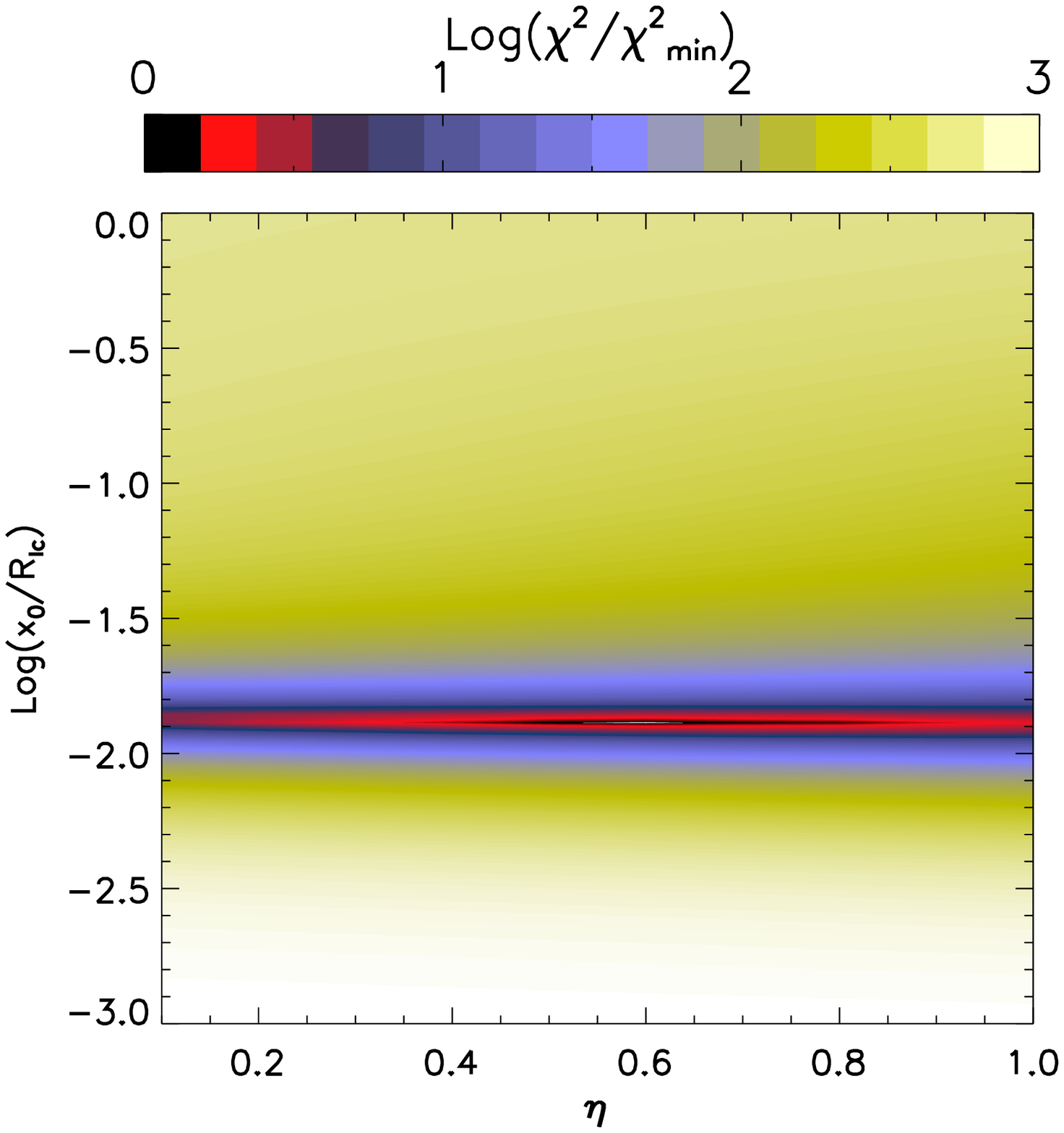}
\includegraphics[width=.32\textwidth]{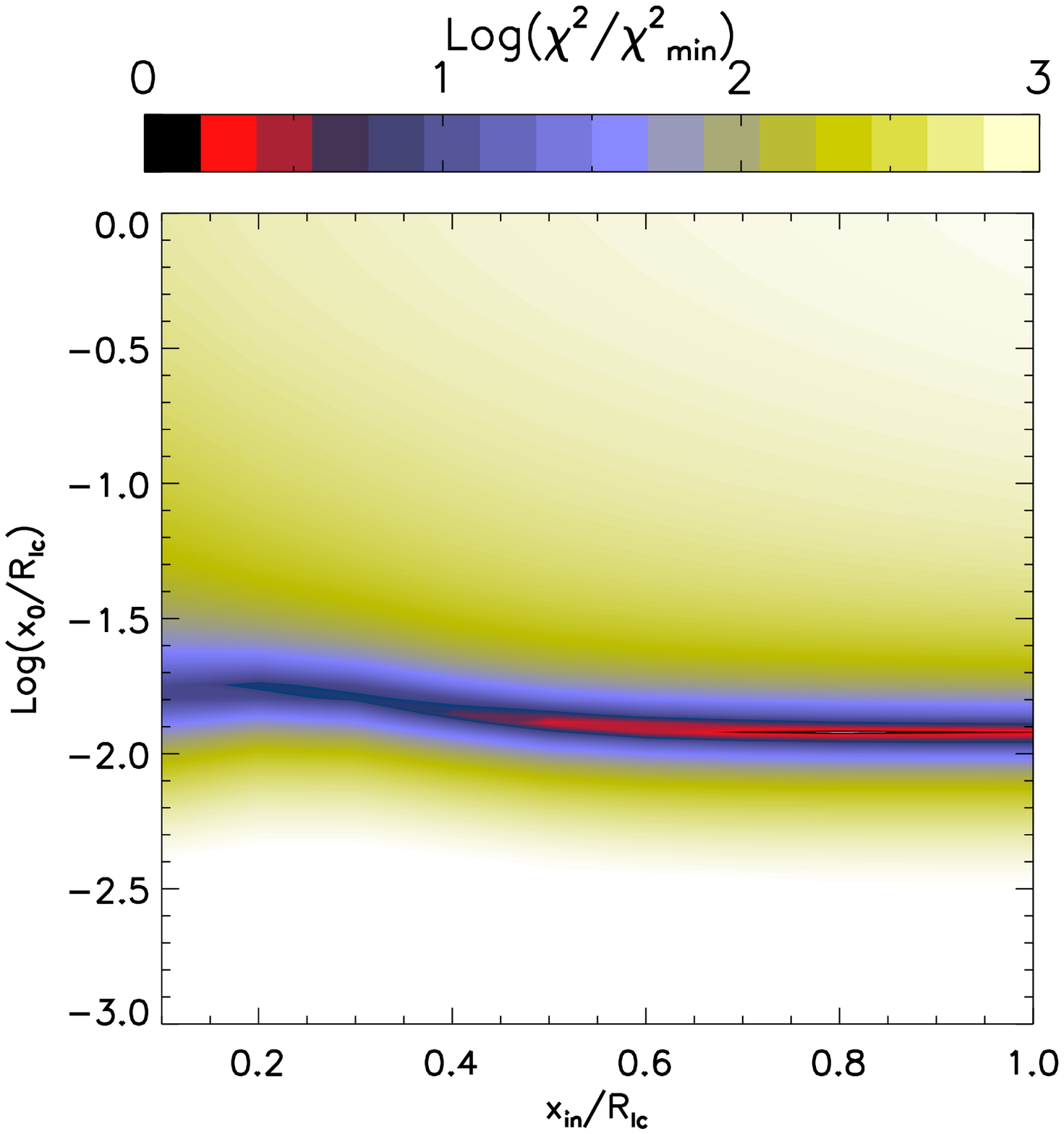}
\caption{Study of the phase-averaged {\em Fermi}-LAT spectrum of the Geminga pulsar. Top left: {\em Fermi}-LAT data and corresponding best-fitting model (see also Table~\ref{tab:pulsars}). Top right: contour plots for the values of $\chi^2/\chi^2_{\rm min}$ in the $\log(E_\parallel)$-$\log(x_0/R_{\rm lc})$ plane, with the values of the remaining parameters given by the baseline gap model, see Table~\ref{tab:parameters}. The white cross represents the best-fitting model of the left-hand panel. Bottom panels: similar contour plots for the planes $b$-$\log(x_0/R_{\rm lc})$, $\eta$-$\log(x_0/R_{\rm lc})$, and $x_{\rm in}$-$\log(x_0/R_{\rm lc})$, respectively, with $\log(E_\parallel {\rm [V~m^{-1}]}) = 7.65$. See text and Appendix~\ref{app:fitting} for details about fitting procedure.}
 \label{fig:gema_av}
\end{figure*}

\section{Results}\label{sec:results}

Instead of the (sub-)exponential cut-off power-law fit, Eq.~(\ref{eq:cutoff}), we aim at fitting the spectra with the expected SC radiation, constraining the physical parameters of our gap model. In order to quantify the relative goodness of the fits, we compare the observed {\em Fermi}-LAT data with the binned theoretical spectra. In Appendix \ref{app:fitting} we give the details about the explored grid of parameters, the re-binning of spectrum, and the definition of the goodness-of-fit indicator $\tilde{\chi}^2$.

According to the model setup described above, for a given pulsar we can fix the values of the spin period $P$, which defines $R_{\rm lc}$, Eq.~(\ref{eq:rlc}), and the surface magnetic field $B_\star$, inferred from Eq.~(\ref{eq:bstar}). Then, we have six parameters that regulate the trajectory: $E_\parallel$, $\eta$, $b$, $x_{\rm in}$, $\Gamma_{\rm in}$, and $\alpha_{\rm in}$, and three parameters describing the effective particle distribution, Eq.~(\ref{eq:distribution}): $N_0$, $x_0/R_{\rm lc}$ and $x_{\rm out}/R_{\rm lc}$.

Since our spectral models are not analytical, and require the computation of the particle trajectory, we cannot perform a systematic, large and blind coverage of the multidimensional space of parameters. Instead, as a first step, we start by identifying which parameters have the largest impact on the spectrum and focus on those, relying also on our previous works \citep{paper0,paper2}.

In Fig.~\ref{fig:trajectories} we saw that the kinematic evolution of particles is very sensitive on the value of $E_\parallel$, while the other five parameters, ($\eta$, $b$, $x_{\rm in}$, $\Gamma_{\rm in}$, $\alpha_{\rm in}$), have non-negligible impact only in the very limited parts of the trajectories. In particular, $b$, $\Gamma_{\rm in}$, and $\alpha_{\rm in}$ will have an impact only if the particle distribution has a width $x_0/R_{\rm lc} \lesssim 10^{-4}$. 

\begin{figure*}
\begin{overpic}[width=0.36\textwidth]{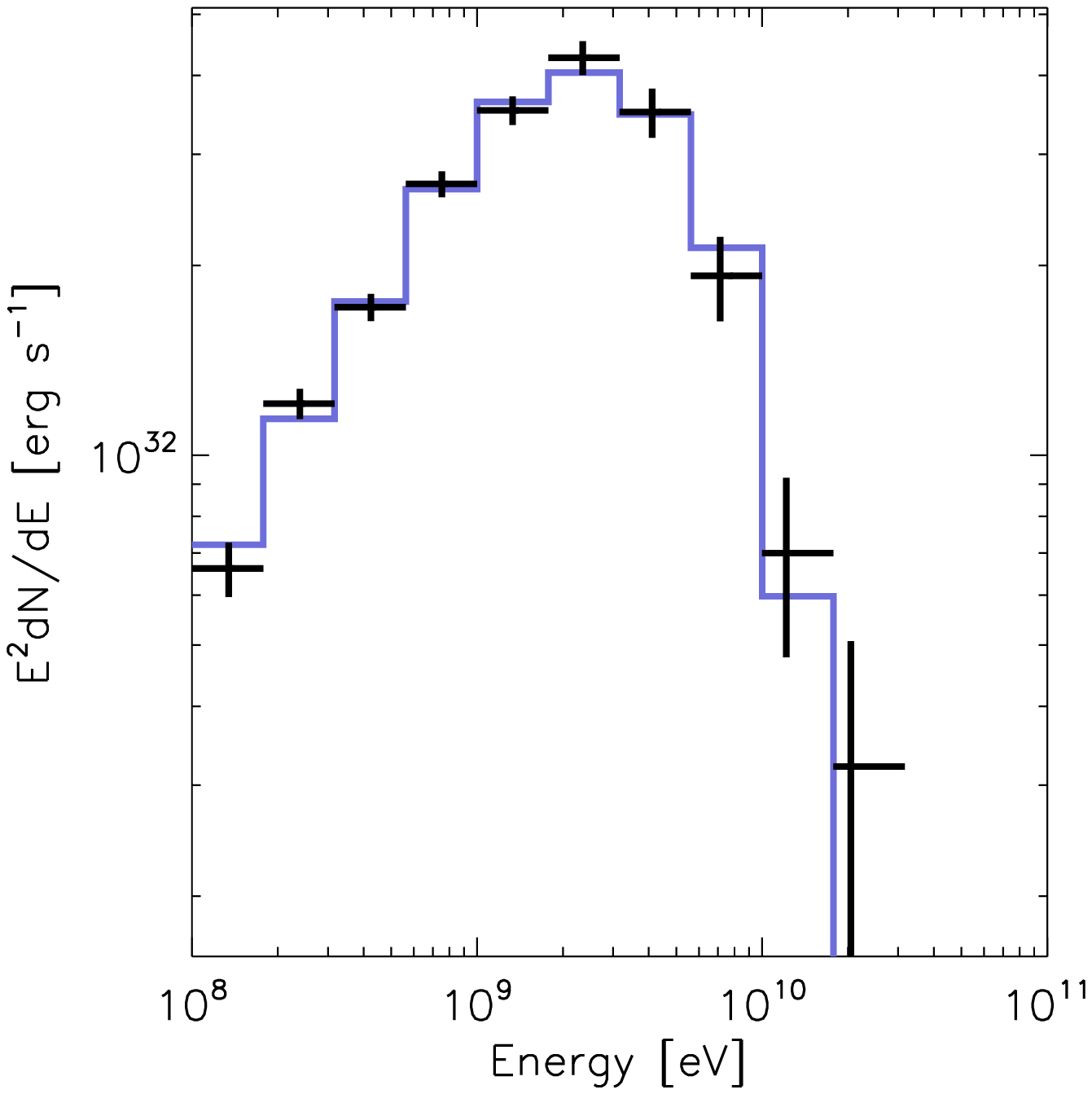}
\put(45,95){Geminga P1}
\end{overpic}
\begin{overpic}[width=0.36\textwidth]{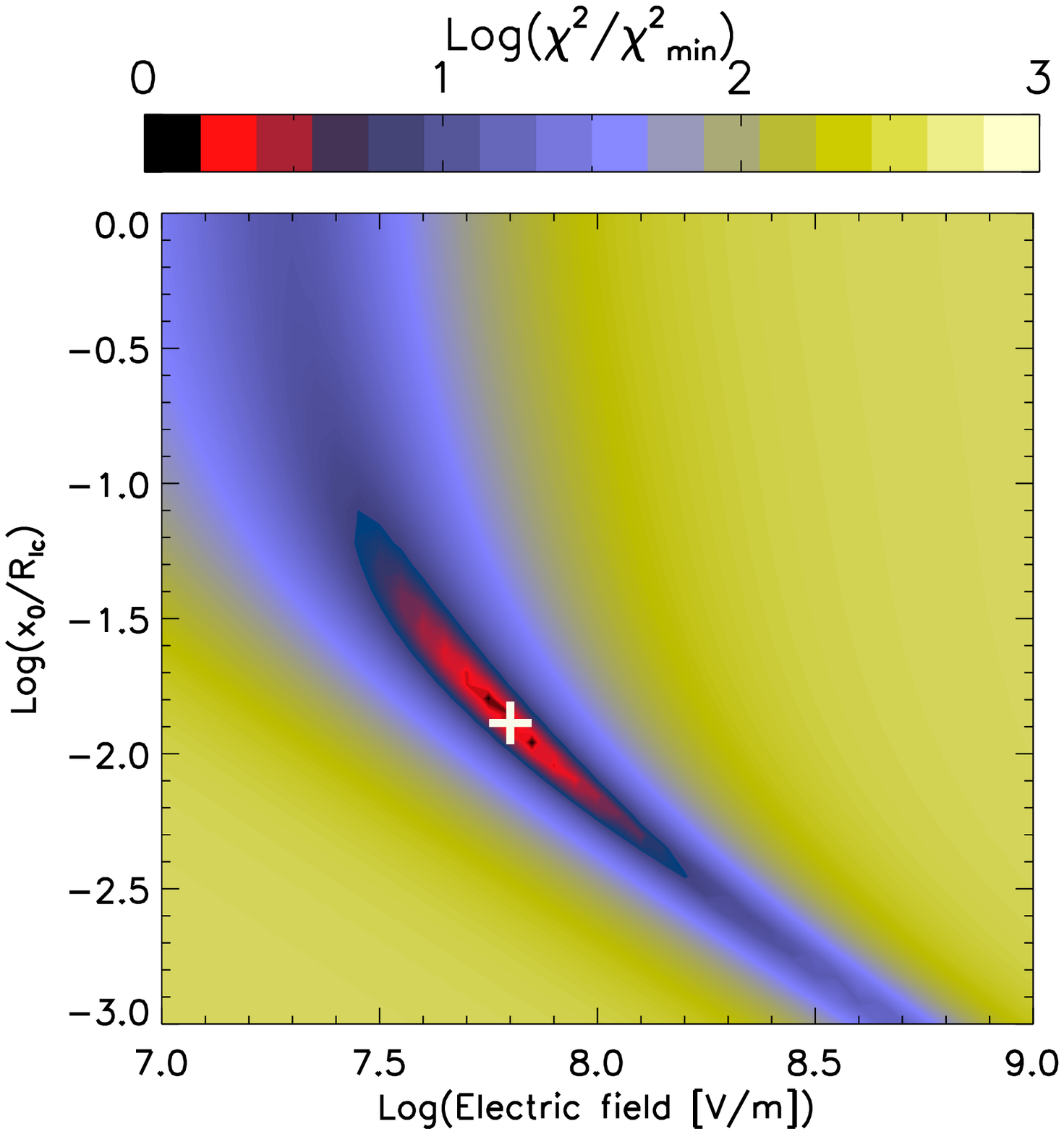}
\put(60,75){Geminga P1}
\end{overpic}
\includegraphics[width=.32\textwidth]{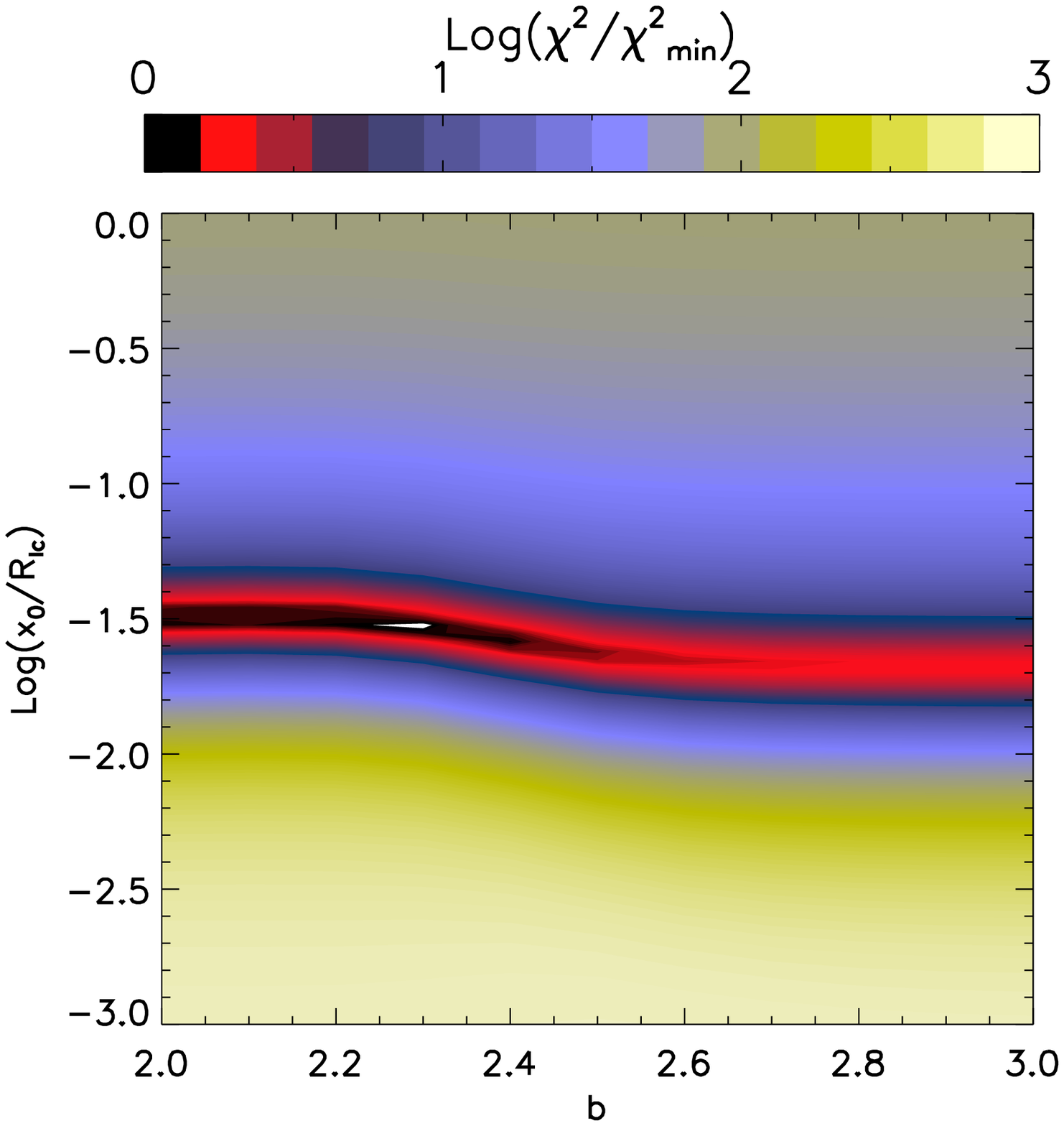}
\includegraphics[width=.32\textwidth]{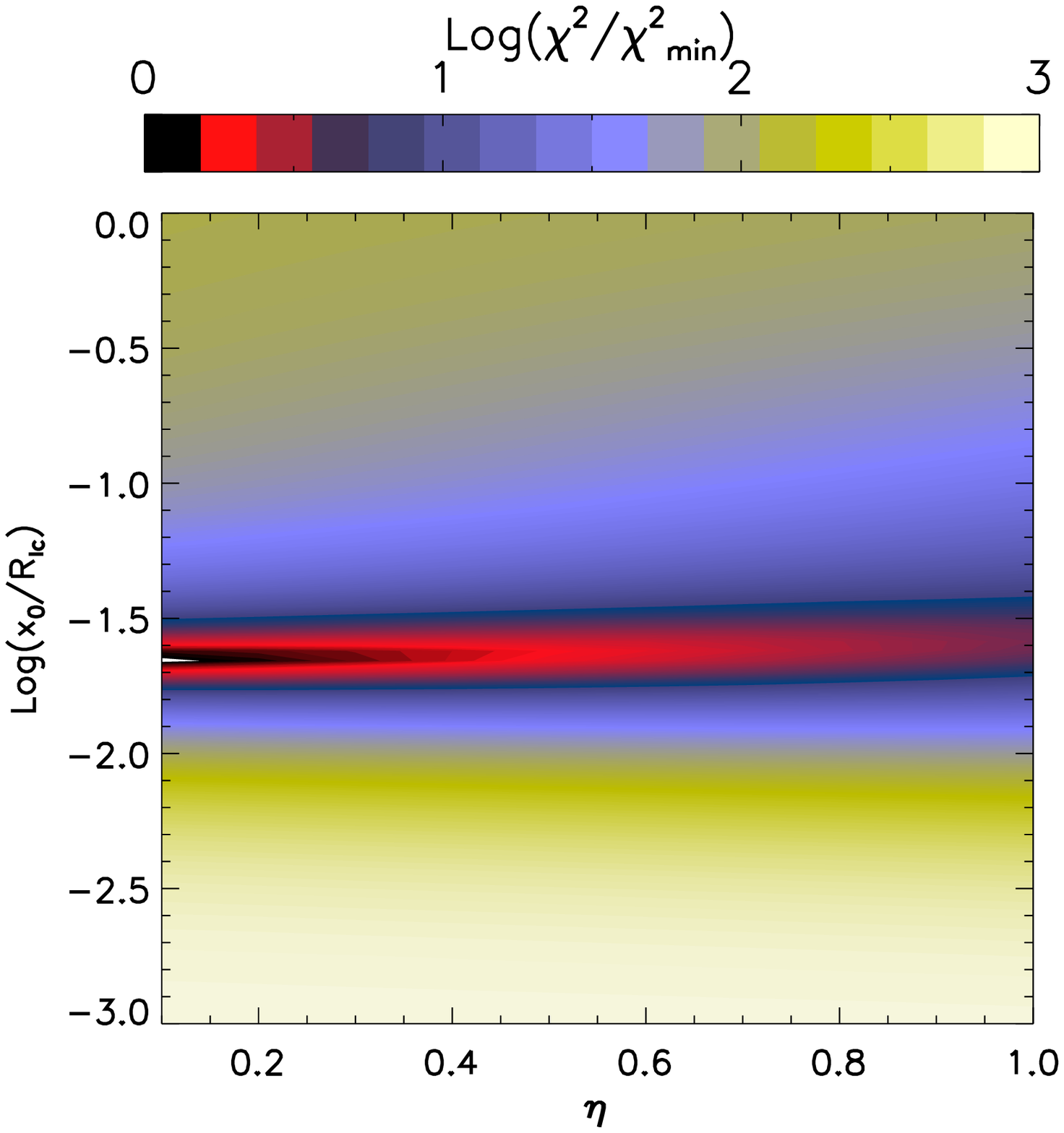}
\includegraphics[width=.32\textwidth]{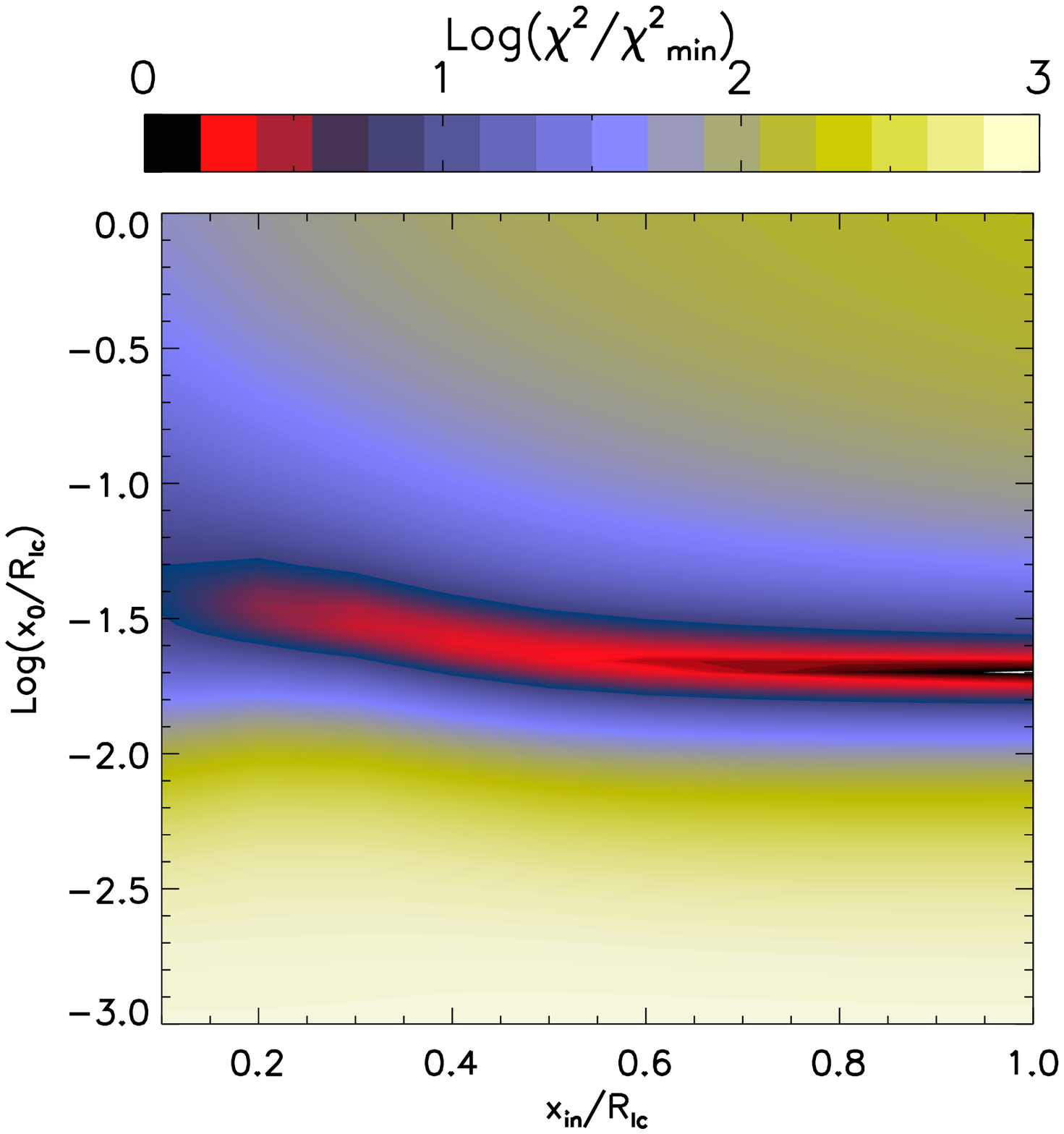}
\caption{Same as Fig.~\ref{fig:gema_av}, but for the phase-resolved spectrum in P1 of the Geminga pulsar.}
 \label{fig:gema_P1}
\end{figure*}

Thus, since the typical photon energy, Eq.~(\ref{eq:echar}), strongly depends on the value of $\Gamma$, $E_\parallel$ is mostly constrained by the energy peaks of observed spectra. The value of $x_0/R_{\rm lc}$ is mostly constrained by the low-energy slope. The value of $x_{\rm out}$ has an effect only if $x_0 \gtrsim x_{\rm out}$, for which the outer part of the gap give important radiation. However, we anticipate here that small values $x_0/R_{\rm lc} \ll 1$ are favored from the spectral fits, thus $x_{\rm out}$ is unconstrained. Lastly, the normalization $N_0$ is picked up by minimizing $\tilde{\chi}^2$, once the other parameters are fixed (see Appendix~\ref{app:fitting}). 

Briefly: the timing-derived parameters, $P$ and $B_\star$, are fixed for a given pulsar, and the three most relevant parameters are: $N_0$, $x_0/R_{\rm lc}$ and $E_\parallel$. In Table~\ref{tab:parameters} we summarize the set of parameters and the treatment we use when we fit data. If not specified, the other parameters are fixed to a baseline model with the values indicated in the third column.

Below, we apply our models to fit data to Geminga, Crab, and Vela pulsars, considering both the phase-averaged and phase-resolved spectra of {\em Fermi}-LAT. In Table~\ref{tab:pulsars} we show their timing properties, the derived rotational energy loss, characteristic age and surface magnetic field, the phases for which we consider the spectra below, appropriate references, and the results of the fit that we illustrate below. Since we focus on the SC radiation, we limit our study to the {\em Fermi}-LAT data.

\subsection{Geminga}\label{sec:geminga}

The Geminga pulsar has been detected in $\gamma$-rays and X-rays. Upper limits on the $E>100$ GeV emission have recently ruled out any hardening above $\gtrsim 50$ GeV \citep{aliu15}. The $\gamma$-ray light curve of the Geminga pulsar is double-peaked, with P1 being the main peak for $E \gtrsim 0.3$ GeV. The second peak, P2, is slightly higher than P1 in the range 0.1-0.3 GeV. For the phase-resolved spectra, we consider the bins associated with the two peaks and the minimum phase, identified according to the fluxes reported by \cite{abdo10c}, and indicated in Table~\ref{tab:pulsars}.

\begin{figure*}
\centering
\begin{overpic}[width=0.36\textwidth]{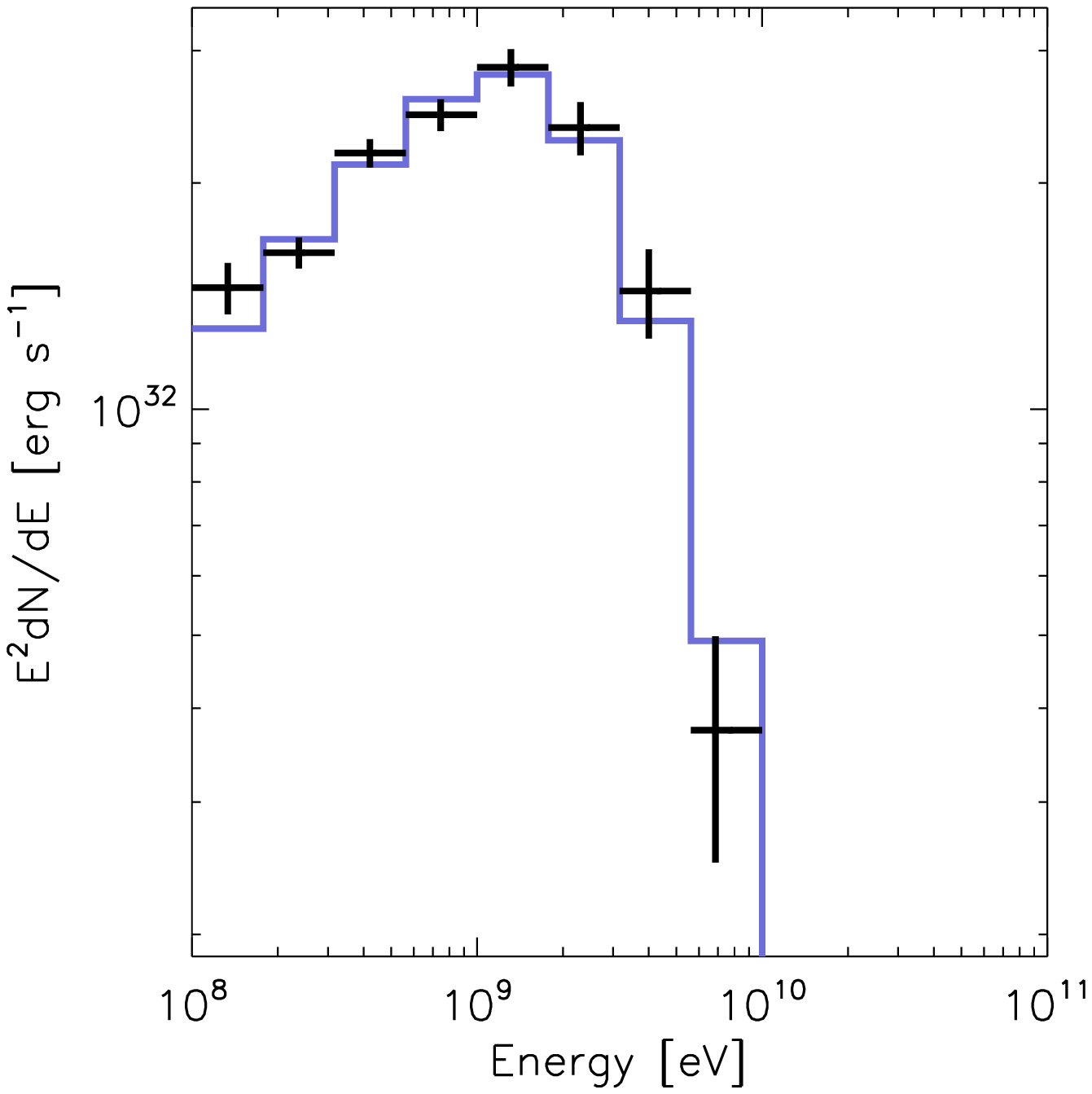}
\put(45,95){Geminga P2}
\end{overpic}
\begin{overpic}[width=0.36\textwidth]{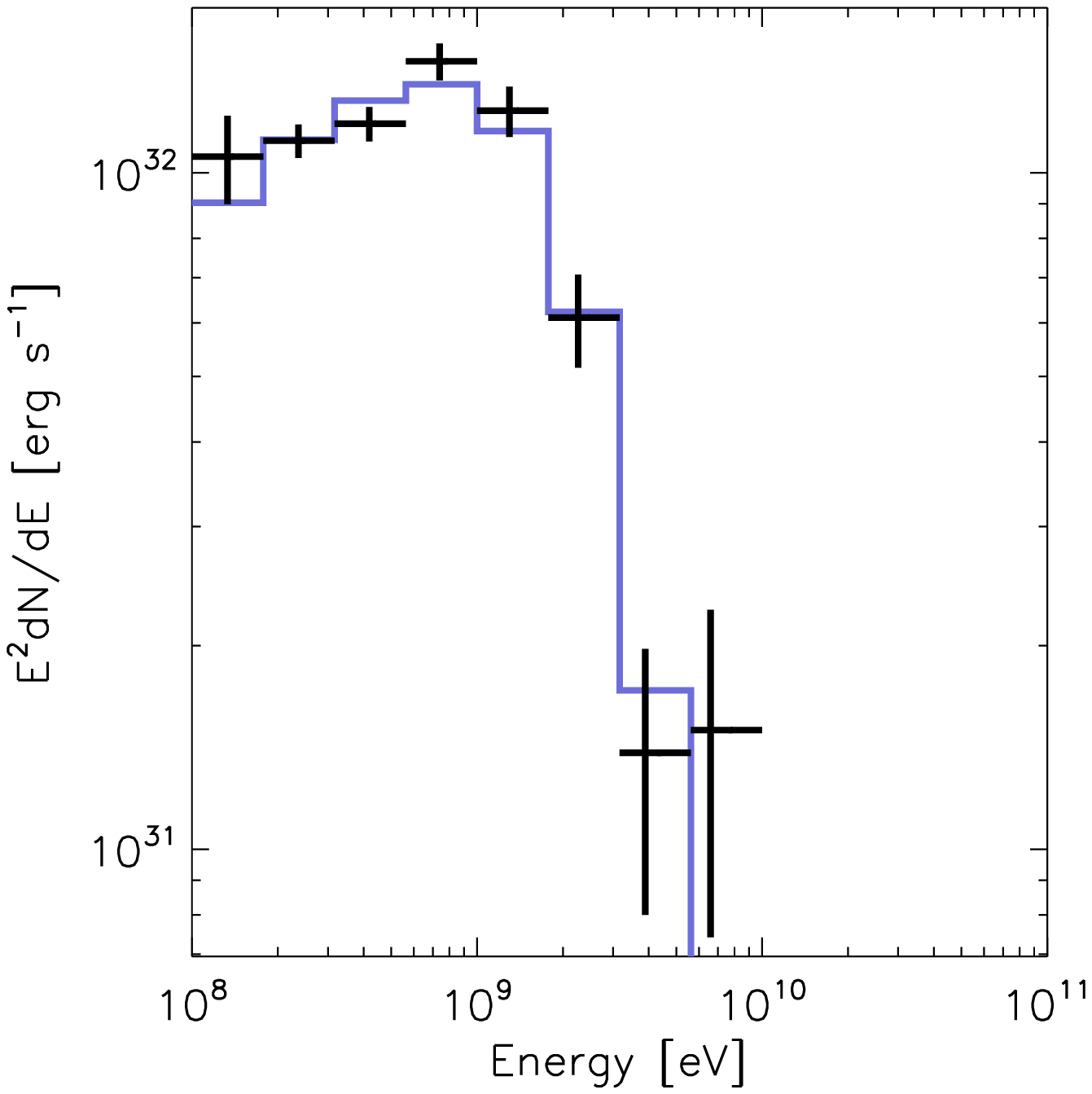}
\put(45,95){Geminga Minimum}
\end{overpic}
\caption{Best-fitting models of the Geminga pulsar for P2 (left) and minimum (right).}
\label{fig:bestfit_gema}
\end{figure*}

We start the discussion with the phase-averaged spectrum. In the top left-hand panel of Fig.~\ref{fig:gema_av} we show the best-fitting model and the observed spectrum, obtained by exploring the $\log(E_\parallel)$-$\log(x_0/R_{\rm lc})$ plane, fitting $N_0$ in each model, and fixing the other parameters to the baseline model of Table~\ref{tab:parameters} (see above and Appendix \ref{app:fitting} for more details about the fitting procedure and grid of explored parameters). The best-fitting values are $E_\parallel =10^{7.65}$ V~$m^{-1}$, $x_0/ R_{\rm lc}\sim 0.013$, and $N_0=1.9\times 10^{31}$ (see also Table~\ref{tab:pulsars}). In the top right-hand panel, we show the related contour plots of $\chi^2/\chi^2_{\rm min}$, where $\chi^2_{\rm min}$ is the minimum best-fitting value (white cross). A low value of $x_0/R_{\rm lc}$ is needed to fit the spectrum below the energy peak, which require that more weight is given to the parts with large $\xi$, i.e., dominated by synchrotron-like emission. This is a clear example of why a uniform distribution, where the radiation would be dominated by the $\xi\ll 1$ part of the trajectory, cannot explain the low-energy slope.

The $\tilde{\chi}^2_{\rm min}$ associated with the best-fitting of the average spectrum is large for three reasons. First, the errors in the low energy part, $\lesssim 1\%$, are very small. Secondly, the high-energy (above 10 GeV) data points are not fitted well. This incompatibility is inherent to the strictly exponential decay of the SC radiation formula, and agrees with the phenomenological sub-exponential cut-off model claimed in \cite{abdo10c}. We come back to this below. Thirdly, further fine-tuning of the explored parameter (e.g., a finer grid of $E_\parallel$ and $x_0/R_{\rm lc}$) would further improve the fit.

Looking at the contour plot, we note a strong anticorrelation between $E_\parallel$ and $x_0/R_{\rm lc}$, with a strip of models having $\tilde{\chi}^2 < 2 \tilde{\chi}^2_{\rm min}$ that constrain $\log(x_0/R_{\rm lc})$ and $\log(E_\parallel)$ within $\sim 20$ and $\sim 5$ per cent, respectively. This anticorrelation, which is seen in all cases we have studied, is interpreted as follows. The low-energy slope with $\mu < 0.25$ can be reproduced only if the initial parts of the trajectory are dominating the observed radiation. A large contribution of such parts, $\xi \gg 1$, can be achieved by different models: the larger the electric field, the shorter is the acceleration length-scale, hence a smaller value of $x_0/R_{\rm lc}$ is needed to consider the parts with relatively low $\Gamma$. In any case, the favored values are $x_0/R_{\rm lc} \ll 1$: we are seeing radiation only from the inner, initial part of the particle trajectories.

As a second step, we check the influence of other parameters. We evaluate the best-fitting spectra for different values of $x_0/R_{\rm lc}$, with $E_\parallel = 10^{7.65}$ V~$m^{-1}$ (best-fitting value), and varying, one by one, the following three parameters: $b \in [2,3]$ (step 0.1, left-hand panel), $\eta \in [0.2,1.0]$ (step 0.1, middle), $x_{\rm in}/R_{\rm lc} \in [0.2,1.0]$ (step 0.1, right-hand panel). Results, drawn in the bottom panels of Fig.~\ref{fig:gema_av}, show that these three parameters are hardly constrainable, apart from apparently favoring $x_{\rm in} \gtrsim 0.5 R_{\rm lc}$ and $b\gtrsim 2.5$. Both weak constrains correspond to models with more efficient radiative losses: low values of $b$ correspond to larger values of $B$ in the outer magnetosphere, low values of $x_{\rm in}$ imply a smaller radius of curvature. In both cases, particles will reach lower values of $\Gamma$. However, the large degeneracy and the much stronger dependence on $E_\parallel$ and $x_0/R_{\rm lc}$ prevents us from taking firm conclusions about these lower limits. The lack of such constraints justifies our initial choice of $E_\parallel$, $x_0/R_{\rm lc}$, and $N_0$ as the only three parameters to be left free in the fit.

Since phase-resolved spectra show important variation between phases, we explore three representatives of them here. In Fig.~\ref{fig:gema_P1} we show the same best-fitting model and contour plots shown for the phase-averaged spectrum, but referred to the spectrum of P1. In these cases, all data can be fitted well by our models, giving a much smaller $\tilde{\chi}^2$ (mainly due to the fewer photons, i.e., larger errors), and fitting well also the spectrum above GeV. The correlations between parameters are similar to the ones of the phase-averaged spectrum, with the only differences in the weak constraints that can be put on the less relevant parameters, $x_{\rm in}$, $\eta$ and $b$. In Fig.~\ref{fig:bestfit_gema} we show the best-fitting models for P2 and minimum phases, which are compatible with observations.

\begin{figure*}
\centering
\begin{overpic}[width=0.36\textwidth]{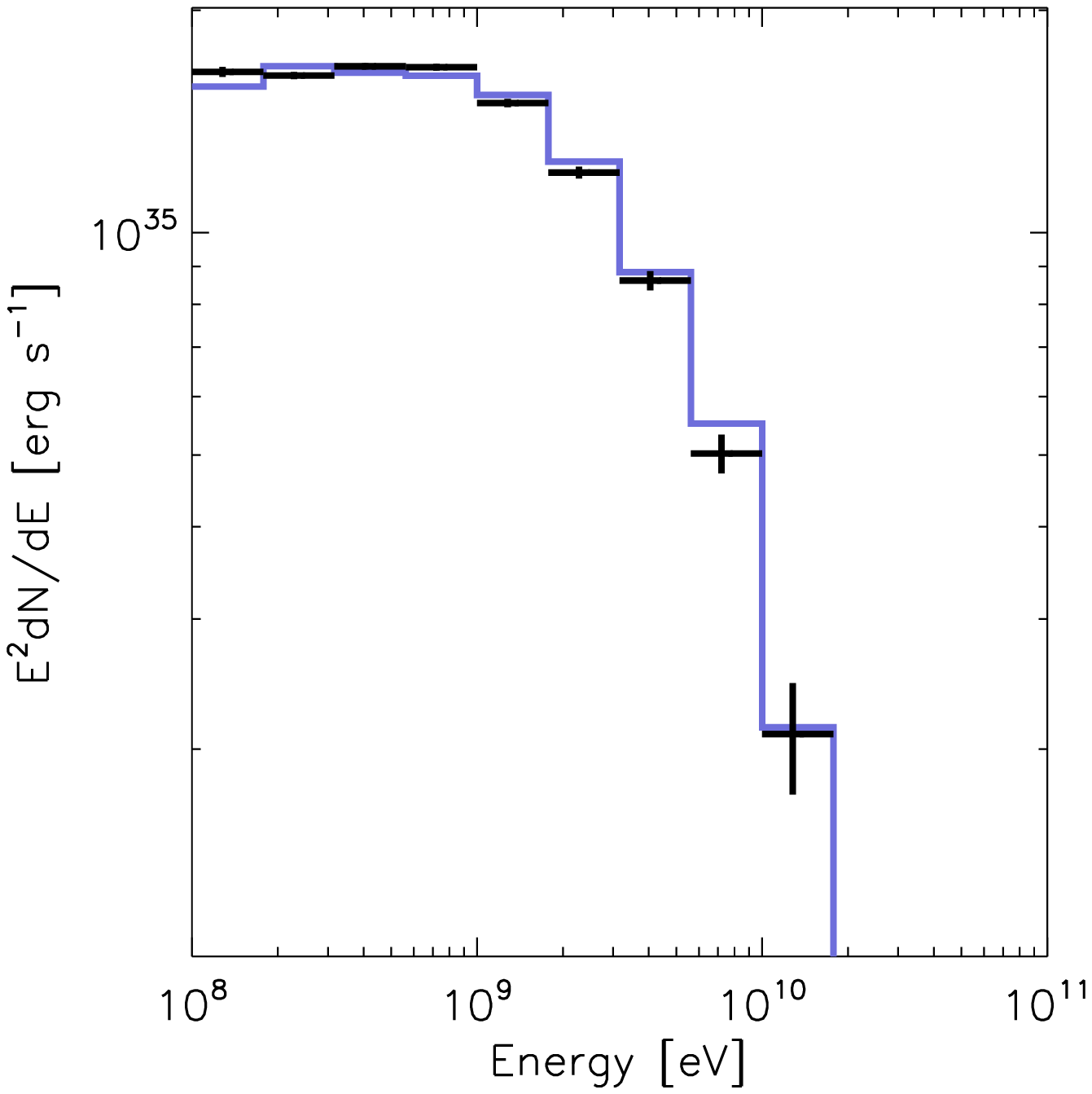}
\put(45,95){Crab average}
\end{overpic}
\begin{overpic}[width=0.36\textwidth]{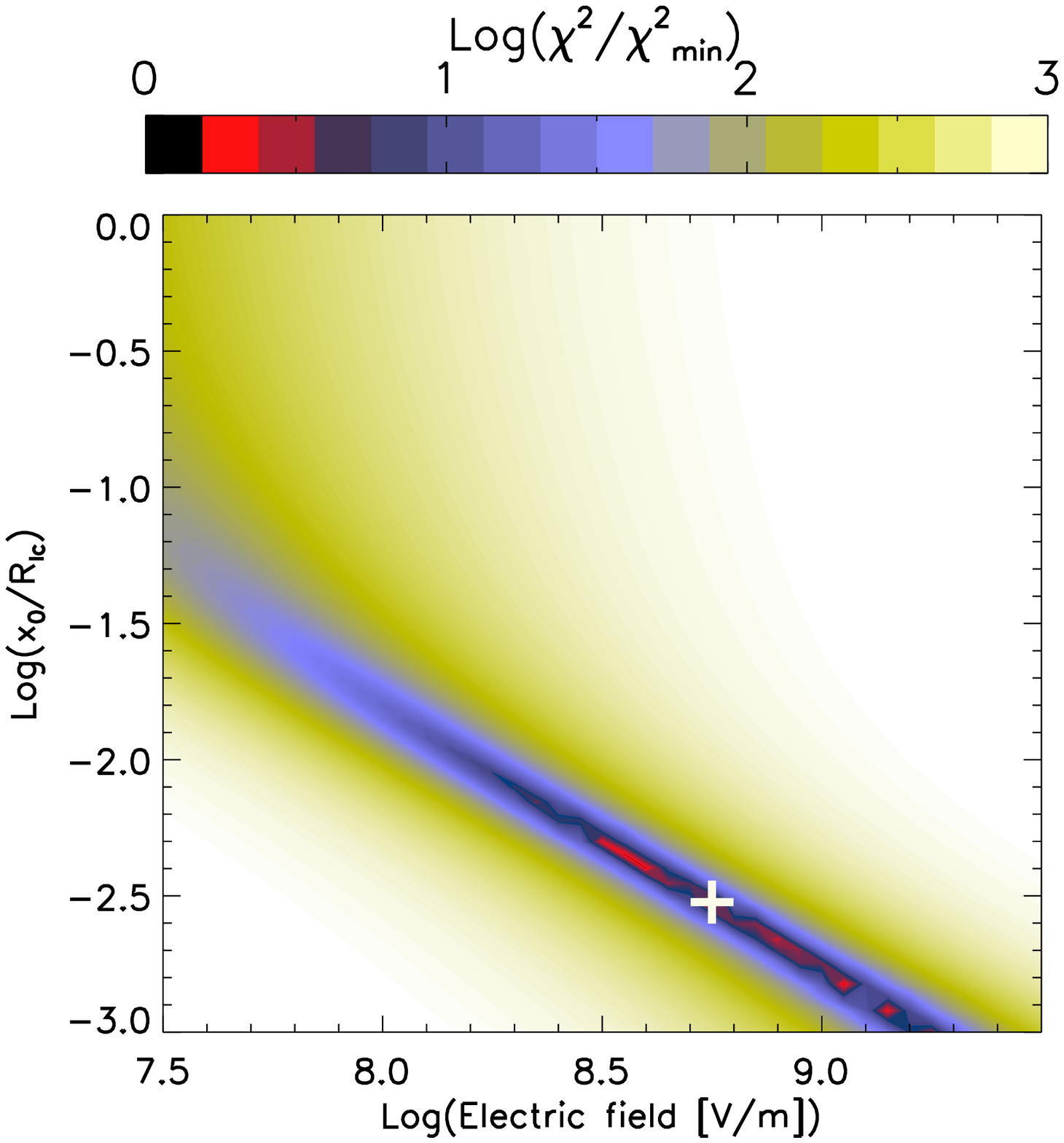}
\put(60,75){Crab average}
\end{overpic}
\caption{Same as top panels of Fig.~\ref{fig:gema_av}, but for the Crab pulsar.}
 \label{fig:crab_av}
\end{figure*}

\begin{figure*}
\centering
\begin{overpic}[width=0.32\textwidth]{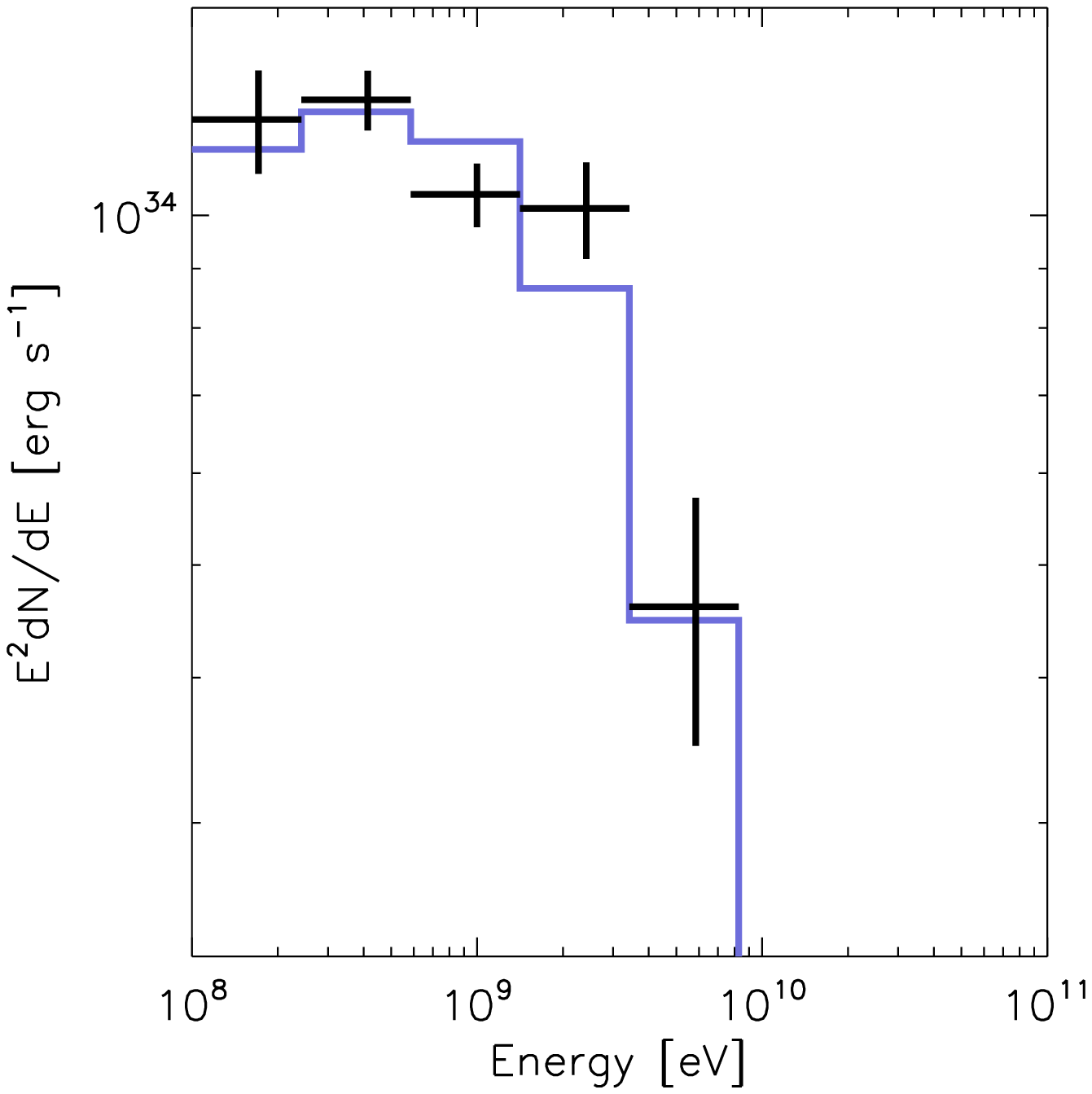}
\put(45,95){Crab P1}
\end{overpic}
\begin{overpic}[width=0.32\textwidth]{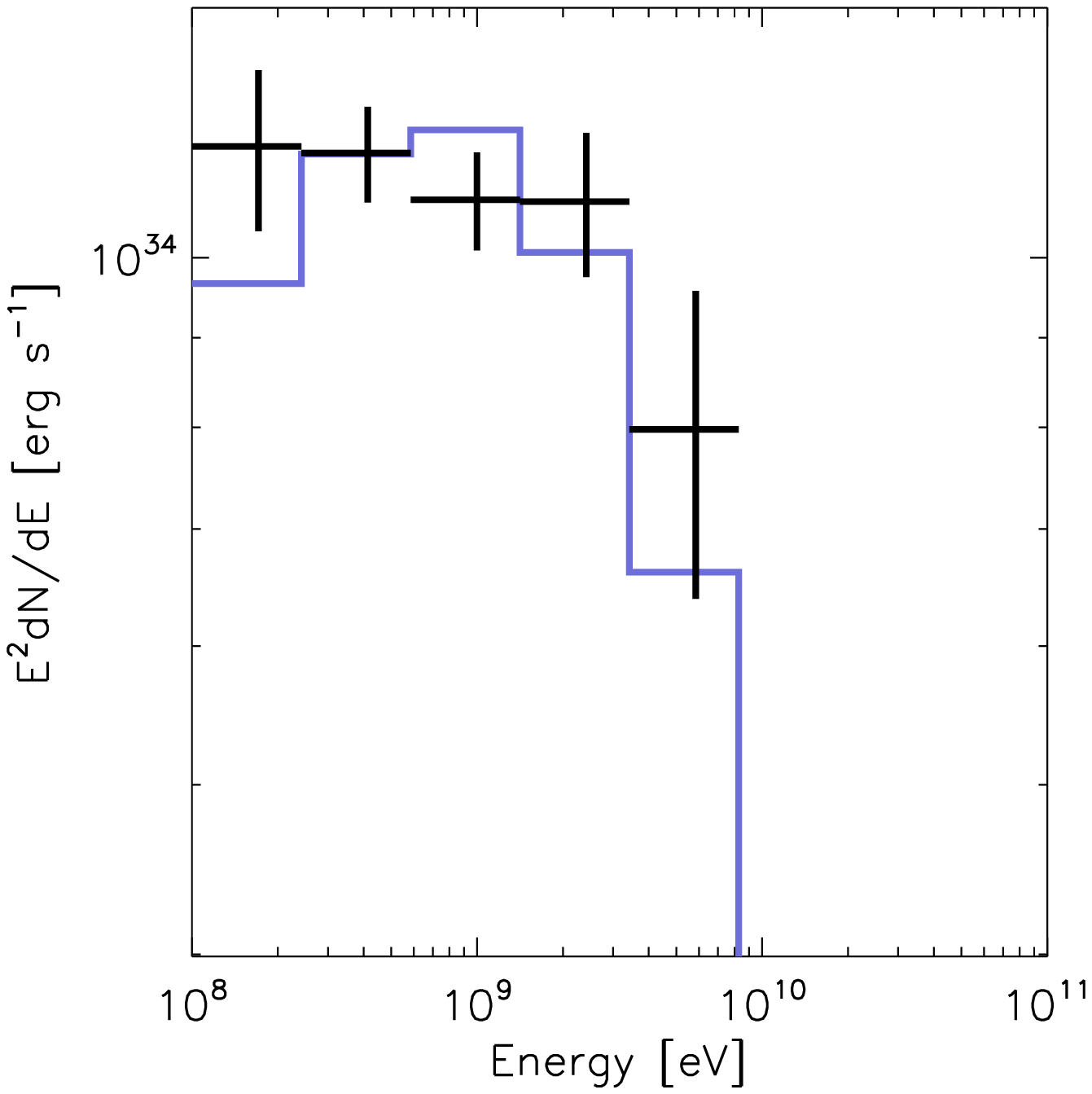}
\put(45,95){Crab P2}
\end{overpic}
\begin{overpic}[width=0.32\textwidth]{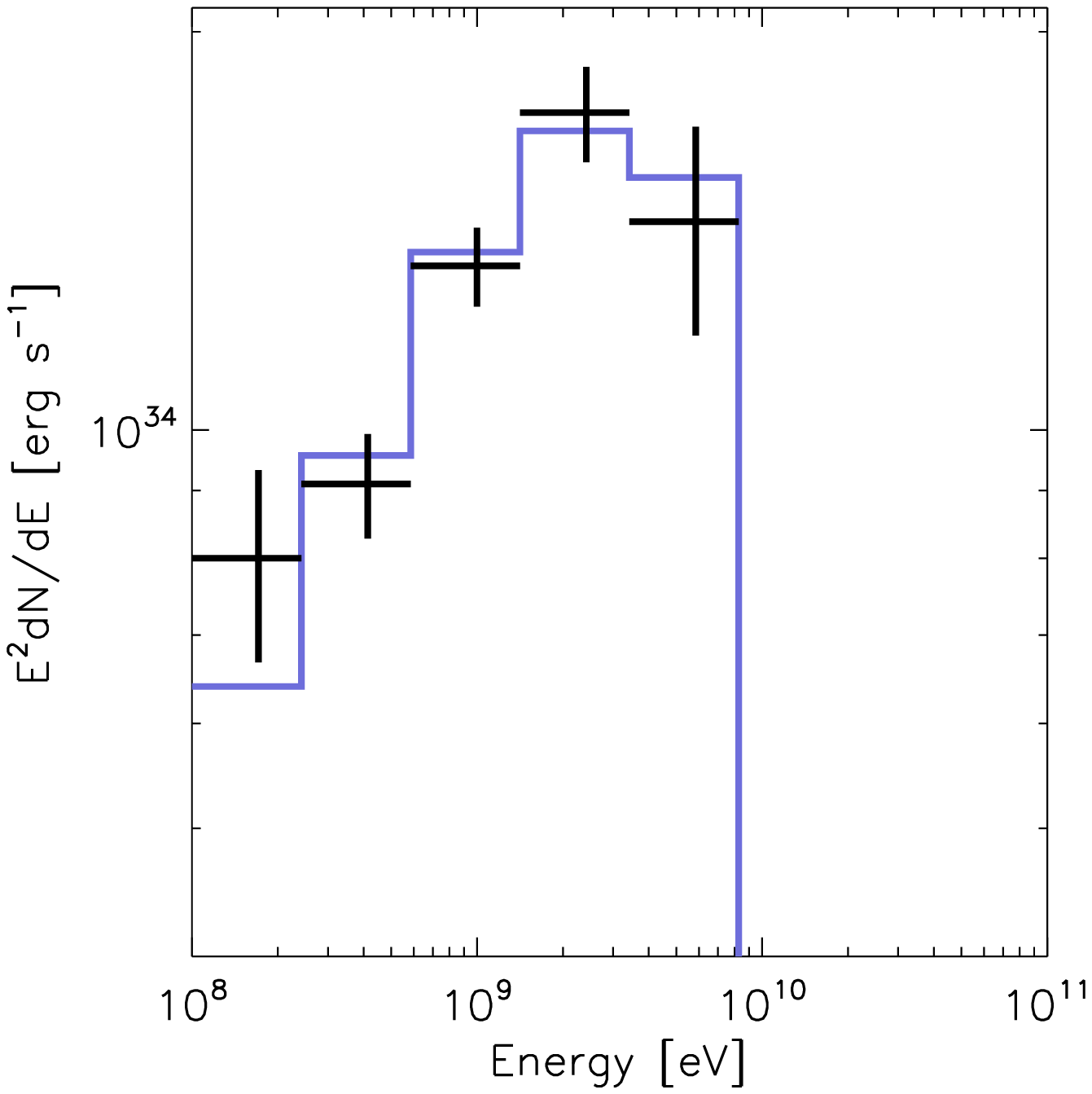}
\put(45,95){Crab Bridge}
\end{overpic}
\caption{Best-fitting models of the phase-resolved spectra of the Crab pulsar in P1 (left), P2 (center), and bridge (right).}
\label{fig:bestfit_crab}
\end{figure*}

In Table~\ref{tab:pulsars} we list the best-fitting parameter values for all cases. Note that in the phases of the peaks the fitted values of $E_\parallel$ and $x_0/R_{\rm lc}$ are similar to those obtained with the phase-averaged spectrum, since the peaks are the main contributors to the average. On the other hand, the minimum shows a much softer spectrum, which can be reproduced by smaller $E_\parallel$ and larger $x_0/R_{\rm lc}$.

In general, we conclude that the high-energy phase-resolved spectra of Geminga are compatible with SC models, if the radiation mostly comes from the loss of perpendicular momentum at the beginning of the trajectories. This agrees with the observational conclusions coming from the detailed phase-resolved analysis reported by \cite{abdo10c}: at each individual phase, the spectrum can be reasonably described by a simple cut-off power-law, with energy peaks, slope indices, and fluxes varying by a factor of a few between different phases. We confirm that the sub-exponential cut-off seen in the phase-averaged spectrum is given by the effect of summing up the spectra coming from different phases. 

\subsection{Crab}\label{sec:crab}

The Crab pulsar has been detected in all energy ranges, from radio up to very high energies. Besides {\em Fermi}-LAT \citep{abdo10a}, pulsed detections have been reported up to 2 TeV by MAGIC\footnote{\url{http://fermi.gsfc.nasa.gov/science/mtgs/symposia/2014/abstracts/185}} \citep{aliu08,aleksic11,aleksic14}, above 100 GeV by VERITAS \citep{aliu11}. At all energy ranges, the light curve presents two well-defined peaks, P1 and P2 of Table~\ref{tab:pulsars}, with almost no lag between different phases. In our analysis, we consider the phase-averaged spectrum, and the phase-resolved spectra for three phase bins, representing P1, P2, and the bridge between them.

As before, we have noted that relatively tight constraints can be put only on $E_\parallel$, $x_0/R_{\rm lc}$, and $N_0$. This is a general feature, so hereafter we will neglect the other parameters of the model. In Fig.~\ref{fig:crab_av} we plot the same contour plot and corresponding best-fitting spectrum to the phase-average data, as we did for the Geminga pulsar, Fig.~\ref{fig:gema_av}. We find a satisfactory fit, with a constrained value of $E_\parallel$ which has to be larger than that in Geminga, in order to reach the observed peak at few GeV. The value of $x_0/R_{\rm lc}$ is very small, again indicating an important contribution by synchrotron-like emission.

In Fig.~\ref{fig:bestfit_crab} we show the best-fitting spectra for P1, P2 and the bridge. The best-fitting values (see Table~\ref{tab:pulsars}) are similar to the phase-averaged spectrum, while the bridge spectrum requires a slightly larger $x_0/R_{\rm lc}$ and lower $E_\parallel$, because of the steeper low-energy slope. In general, the fit to phase-resolved spectra is satisfactory.

Although we are able to reproduce the {\em Fermi}-LAT spectra, we note that the recent detection of the pulsed emission in TeV (not considered here) rules out the SC model to be the only responsible mechanism, indicating a likely important contribution of the IC mechanism at energies $\gtrsim 10$ GeV.

\begin{figure*}
\centering
\begin{overpic}[width=0.36\textwidth]{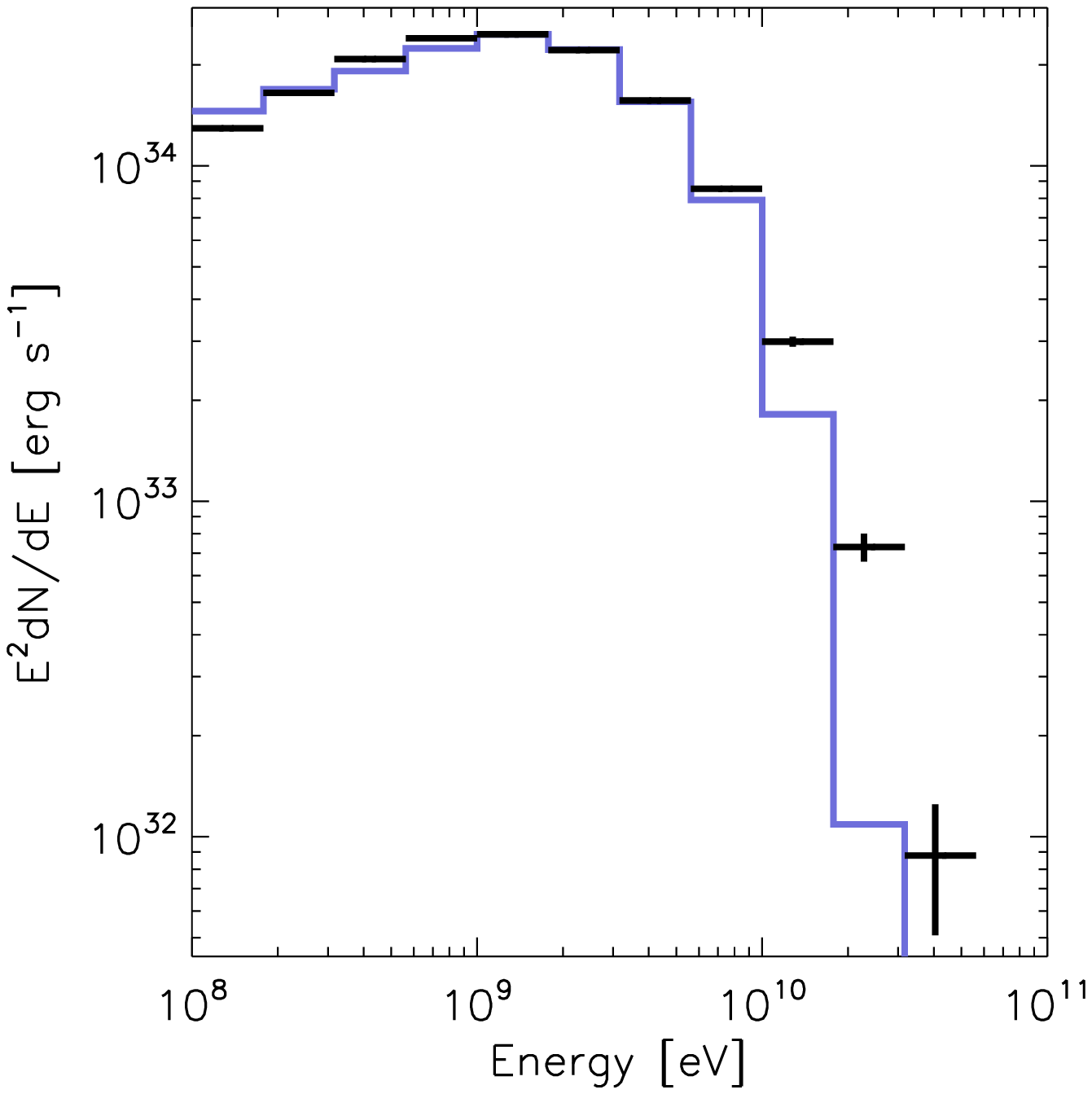}
\put(45,95){Vela average}
\end{overpic}
\begin{overpic}[width=0.36\textwidth]{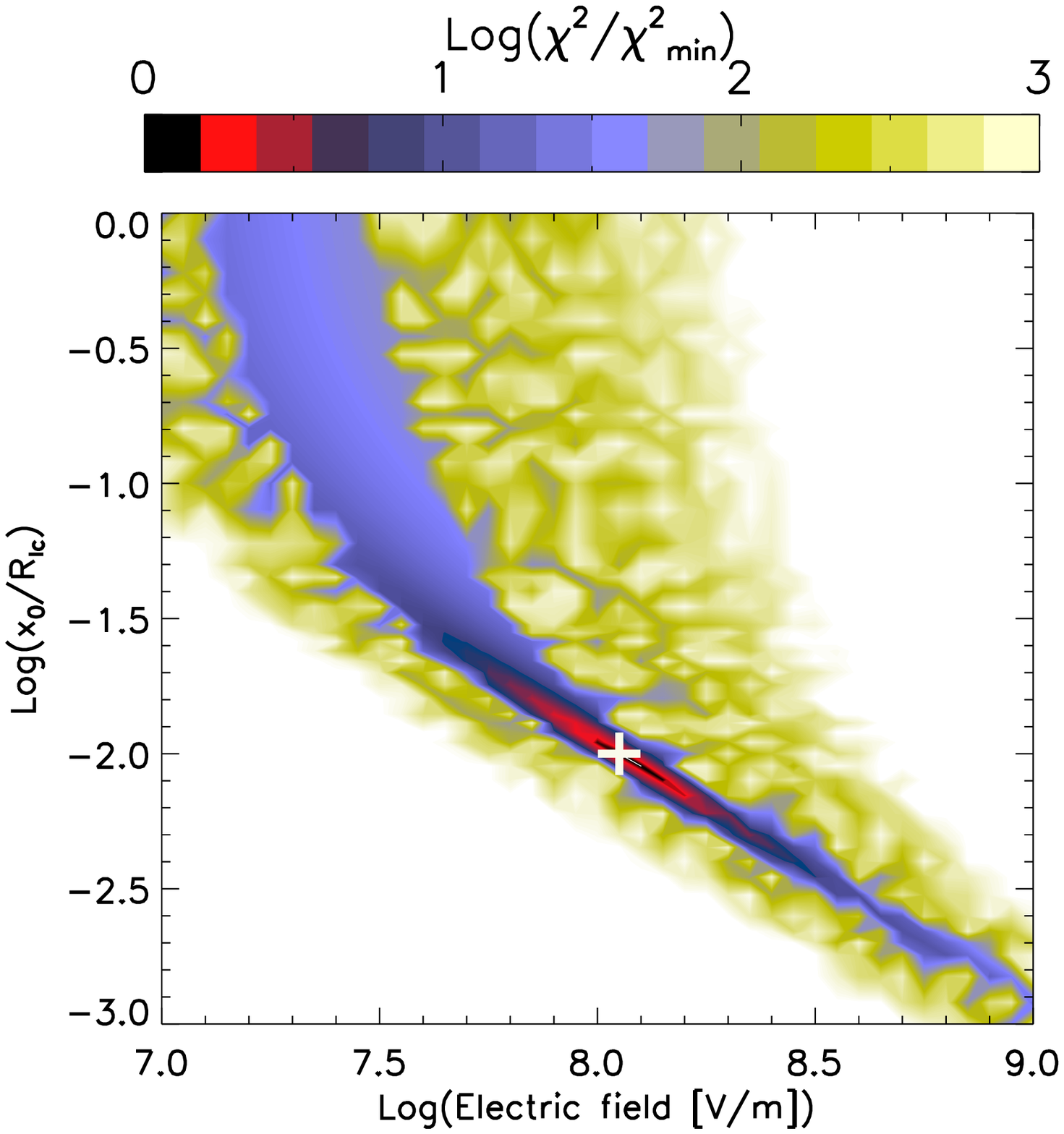}
\put(60,75){Vela average}
\end{overpic}
\caption{Same as top panels of Fig.~\ref{fig:gema_av}, but for the Vela pulsar.}
 \label{fig:vela_av}
\end{figure*}

\begin{figure*}
\centering
\begin{overpic}[width=0.32\textwidth]{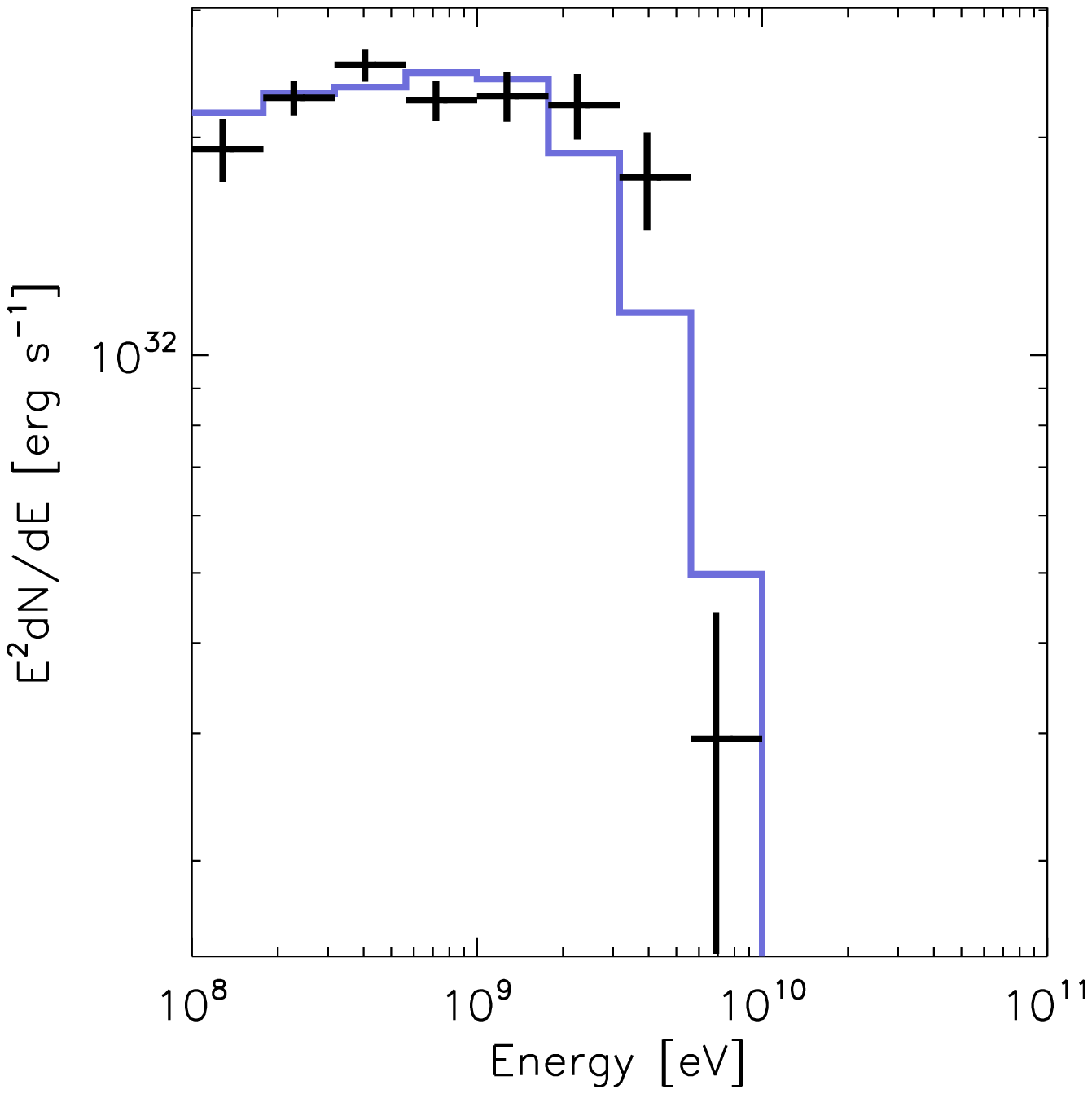}
\put(45,95){Vela P1}
\end{overpic}
\begin{overpic}[width=0.32\textwidth]{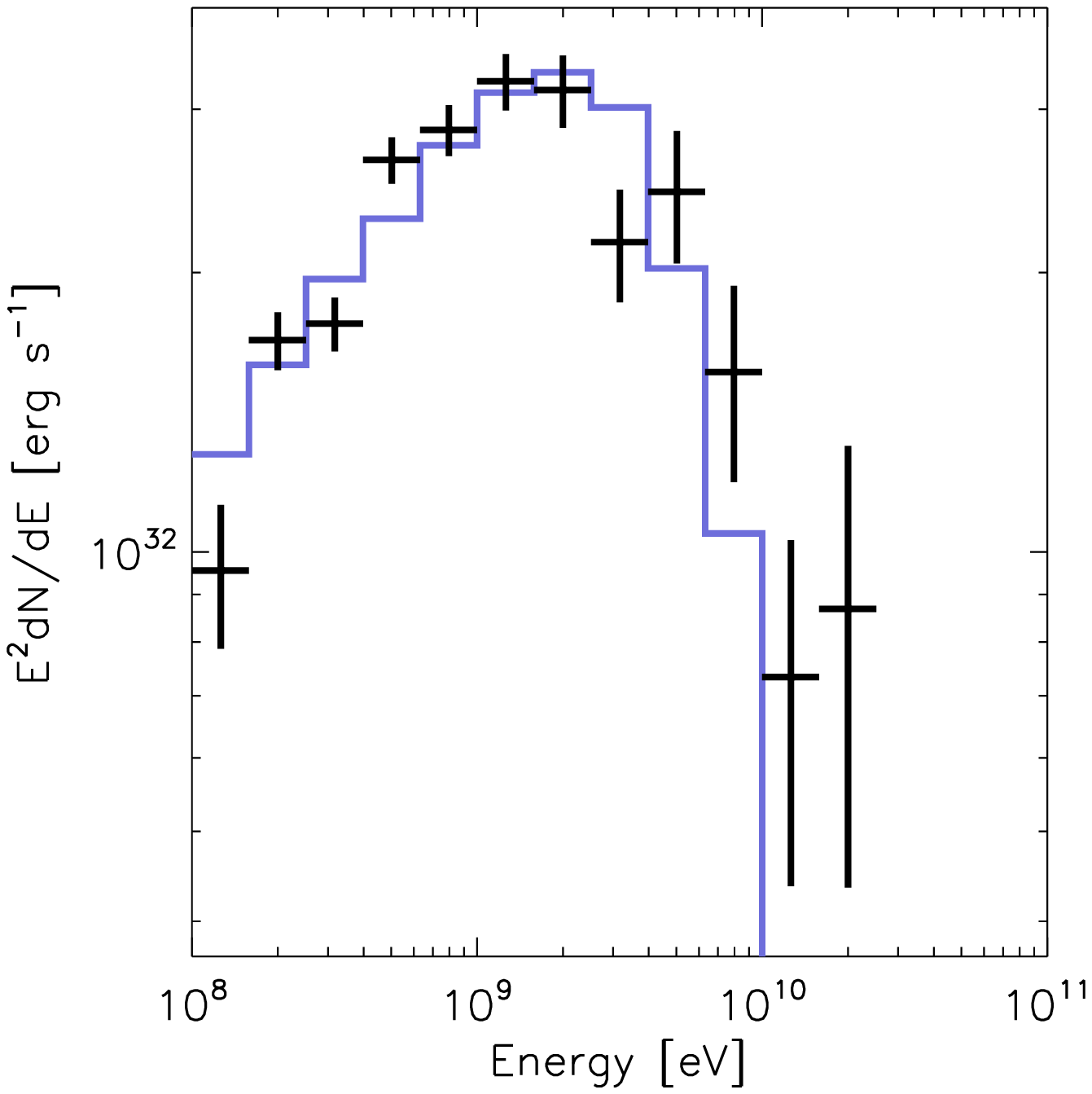}
\put(45,95){Vela P2}
\end{overpic}
\begin{overpic}[width=0.32\textwidth]{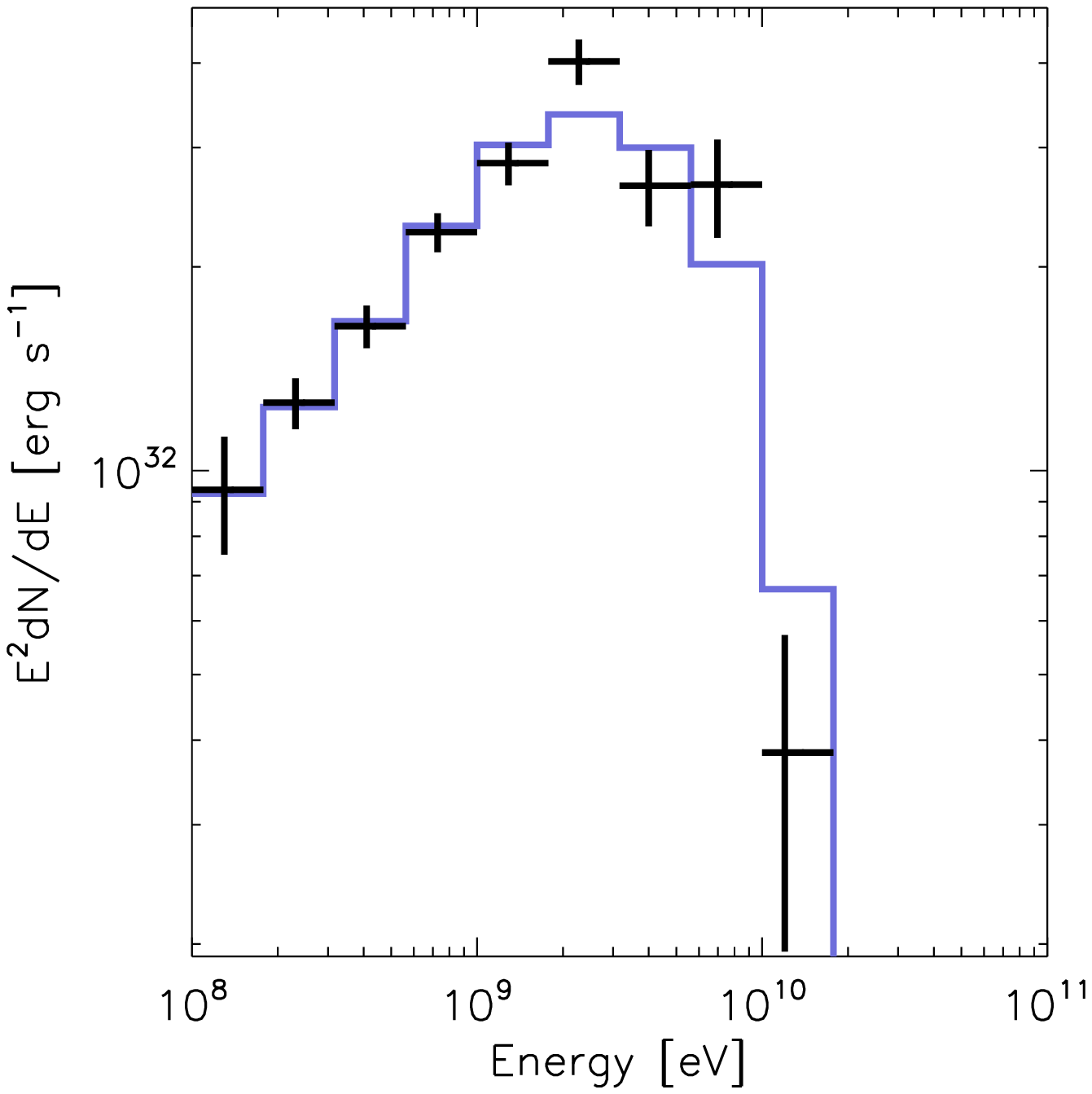}
\put(45,95){Vela P3}
\end{overpic}
\caption{Best-fitting models of the phase-resolved spectra of the Vela pulsar in P1 (left), P2 (center) and P3 (right).}
 \label{fig:bestfit_vela}
\end{figure*}

\subsection{Vela}\label{sec:vela}

The Vela pulsar has been detected from radio to very high energies. Pulsed emission has been detected above 50 GeV (\citealt{leung14}, recently also by the H.E.S.S. telescope\footnote{\url{http://www.mpg.de/8287998/velar-pulsar}}), similarly to the Crab pulsar. It presents a large variability of light curves depending on the energy. In order to study the phase-resolved $\gamma$-ray spectra, we consider data coming from the three phases listed in Table~\ref{tab:pulsars}, corresponding to the two peaks, P1 and P2, which are the maxima below and above 1 GeV, respectively, and an inter-pulse phase, P3, corresponding to a third peak visible at $E \gtrsim 3$ GeV.

In Fig.~\ref{fig:vela_av} we show the contour plot of $\chi^2/\chi^2_{\rm min}$, qualitatively similar to the previous cases, and the corresponding best-fitting spectrum for the phase-averaged spectrum and its best fit for the Vela pulsar.  As before, the tail at hard energy cannot be reproduced by SC models. The best-fitting value of $E_\parallel$ is intermediate between the Crab and Geminga cases, and again we need $x_0/R_{\rm lc}\ll 1$.

In Fig.~\ref{fig:bestfit_vela} we show the phase-resolved spectra for the three phases, and the best-fitting obtained within our models.\footnote{The values are not the same as reported in the Fig.~10 of \cite{abdo10b}, in which the flux values are renormalized by the inverse of the phase bin width (T. Johnson, private communication).} The spectra in P1 and P3 are compatible with the SC emission model, while in P2 the residuals at high energies could point to contributions from IC scattering.

\section{Conclusions}\label{sec:conclusions}

In this work, we have applied a SC model to the phase-averaged and phase-resolved {\em Fermi}-LAT spectra of the three brightest and best-studied $\gamma$-ray pulsars: Geminga, Crab, and Vela. The theoretical spectra rely on a 1D gap model based on effective parameters, whose exploration allow us to systematically fit data. For a given pulsar, $P$ and $B_\star$ are fixed by the observed timing properties. These define $R_{\rm lc}$, Eq.~(\ref{eq:rlc}), and the surface magnetic field $B_\star$,  Eq.~(\ref{eq:bstar}). Then, the properties of the emitting region can be effectively described by the following effective parameters: the accelerating electric field $E_\parallel$, four geometrical parameters, $\eta$, $b$, $x_{\rm in}/R_{\rm lc}$, and $x_{\rm out}/R_{\rm lc}$, two kinematic parameters, $\Gamma_{\rm in}$ and $\alpha_{\rm in}$, and two parameters describing the effective particle distribution: $N_0$ and $x_0/R_{\rm lc}$.

A main conclusion is that $E_\parallel$, $x_0/R_{\rm lc}$, and $N_0$ have much more influence on spectra than the remaining six parameters. Thus, observational data can quantitatively fitted by our model varying just these three parameters (which are, in number, one parameter less than the sub-exponential cut-off power-law employed by {\it Fermi}-LAT to describe the data), keeping the others fixed. $E_\parallel$ is mainly constrained by the energy of the peak, $x_0/R_{\rm lc}$ by the low-energy slope, and $N_0$ by the flux. 

\begin{figure}
\centering
\includegraphics[width=.45\textwidth]{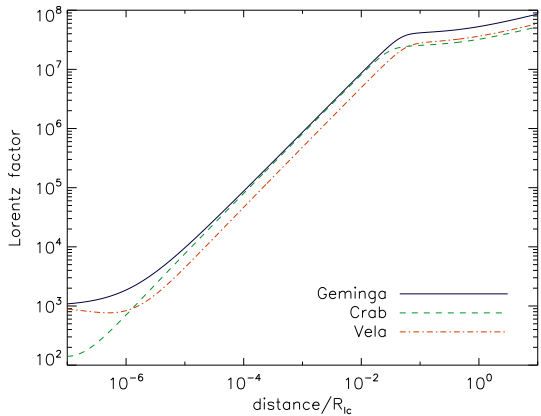}
\caption{Lorentz factors versus normalized distance along the gap, $x/R_{\rm lc}$, for the best-fitting models of the phase-averaged spectra of the three pulsars.}
 \label{fig:gamma_bestfits}
\end{figure}

In Fig.~\ref{fig:gamma_bestfits} we show the evolution of the Lorentz factor for the best-fitting models of the three phase-averaged pulsars. All curves of $\Gamma(x/R_{\rm lc})$ versus normalized distance along the line are similar, reaching similar values of $\Gamma \sim$ few $10^7$ at a distance $x \sim 10^{-2}-10^{-1} R_{\rm lc}$, a length-scale comparable with the inferred values of $x_0$.

In all cases, the best-fitting models are consistent with radiation mostly coming from the initial part of the particle trajectories ($x_0/R_{\rm lc} \lesssim 10^{-2}$), where the perpendicular momentum losses (i.e., synchrotron losses) are not negligible ($\xi \gtrsim 1$, see Fig.~\ref{fig:trajectories}). This constraint is related to the relatively flat slope of the low-energy part, common to many other pulsars, and represents strong, qualitative indications about physical processes. In the framework of a SC interpretation of the pulsed emission, this can be related to several possible physical effects. The beaming angle of the emitted radiation is inversely proportional to $\Gamma$, thus low-energy photons, mostly coming from low-$\Gamma$ particles, are easier to be detected. Moreover, the cascade process, not simulated in our effective approach, could actually provide a large number of particles quickly losing perpendicular momentum in synchrotron-like radiation.

% We find that the position of the gap is poorly constrainable, thus it is difficult to set the height at which radiation is produced. This would be an interesting cross-check of the constraints on the gap position coming from the study of the light curves.  The small region in which the acceleration takes place in our models (of the order of $x_0/R_{\rm lc}$) is consistent with having almost no lag between the pulse profile seen at different energies in the Crab pulsar. As a matter of fact, most of the {\em Fermi}-LAT radiation would originate from particles with relatively low Lorentz factors. As they are accelerated to their saturated values, they can boost X-ray or background photons up to the observed energies by IC. Since the acceleration length-scale is much less than the light cylinder, i.e., the locations of the emitted radiation are close, the lag between energy range is negligible. Another qualitative feature is that the pulse width gets smaller with the energy, and this can be explained by the fact that the beam is narrower at larger energies \citep{aliu11}.

%Such values are at least not in contrast with the recent study of the acceleration of bulk motion of plasma due to magneto-centrifugal mechanism in the magnetosphere \citep{bogovalov14}. 

% f resulting from eq. 60 of Paper I:  Crab 0.0696647  Vela 0.154513  Geminga 0.811975
% N_0/(B12/P) gives (a. units):   0.331681     0.250503      1.16043

For Geminga and Vela pulsars, our best fits to the phase-averaged spectra show residuals after the $\sim$ GeV peak. However, most of the studied phase-resolved spectra, at each individual phase, are compatible with purely SC radiation. In particular, the resulting sub-exponential cutoff in the phase-averaged spectrum of Geminga is coming from the conjunction of individual exponentially decaying spectra showing different cutoff energies. As a consequence, it is natural not to expect any detection from Geminga above tens of GeV, in agreement with VERITAS and MAGIC observations \citep{aliu15}.

In some other cases, such as in at least one of the phases of the Vela pulsar, a sub-exponential cut-off results in important residuals at high energies even in the phase-resolved measurements. These results are consistent with the reported phenomenological fits by {\it Fermi}-LAT \citep{abdo10a,abdo10b,abdo10c}. Another qualitative difference among the three pulsars considered is that Vela and Crab pulsars show pulsed emission at several tens of GeV, and up to 2 TeV, respectively. Detections at such high energies need to be explained by IC scattering. \cite{leung14} proposed a spectral model for Vela based on curvature radiation, consisting in the superposition of different outer gap structures, leading to different energy peaks. However, their model fails to reproduce the low-energy slope, and would be increasingly in tension the higher the energy at which pulsations are detected.

Note that even such mismatch observed at large energies in some cases does not invalidate the SC as the main mechanism of radiation below $\sim$GeV. Instead, it likely points to extra-contributions coming from IC, dominating at large energies only in a few cases. Recent stacked searches for pulsations by \cite{mccann14} indicate no strong average emission at $E\gtrsim 50$ GeV, which could make the SC model valid for most pulsars, with Crab and Vela pulsars being exceptions. Finally, note that the three pulsars span a range of two orders of magnitude in the estimated age (see Table~\ref{tab:pulsars}). The rotational energy differences could explain the different importance of the IC mechanism, which could be more effective for young, more energetic and hotter NSs, due to the larger X-ray flux from the surface.

Comparing the obtained best-fitting values for the three pulsars, we note that the Crab pulsar needs larger values of $E_\parallel$ and $N_0$, compared with Vela and Geminga, for which the best-fitting values are similar. The similar values of $N_0$ between Vela and Geminga, in particular, are in contrast with the large difference (a factor $\sim 200$) between their rotational energies, and reflects the high efficiency of Geminga.

A complementary study to our effective approach should ideally include 3D simulations of particle dynamics, considering the interaction with the radiation, leading also to the IC contribution estimate, with all the geometrical and beaming effects considered (and probably fitting on the gap location). Moreover, the inclusion of multiwavelength data and the study of the light curves should better constrain the models.

In a forthcoming work, we will apply our models to systematically study the entire population of $\gamma$-ray pulsars having good-quality data. Thus, we will constrain the values of $E_\parallel$, $x_0/R_{\rm lc}$, and $N_0$, and look for trends in the population. Since good-quality phase-resolved spectra are available only for a minority of pulsars, one can work with the best phase-averaged spectra, supported by the fact that, for the three cases here studied here and for which phase-resolved analysis exist, the obtained best-fitting values of $E_\parallel$ and $x_0/R_{\rm lc}$ are similar to the phase-resolved spectra at the peaks (see Table~\ref{tab:pulsars}). 

%%%%%%%%%%%%%%%%%%%%%%%%%%%%%%%%%%%%%%%%%
\section*{Acknowledgements}
%%%%%%%%%%%%%%%%%%%%%%%%%%%%%%%%%%%%%%%%%

This research was supported by the grant AYA2012-39303 and SGR2014-1073. We thank the referee for the valuable comments.

\bibliography{og}

\appendix

\section{Fitting procedure}\label{app:fitting}

The aim of the fitting procedure is to constrain the three relevant parameters of our model, $E_\parallel, x_0/R_{\rm lc}, N_0$, by comparing the theoretical and observational spectra, the latter taken from the publicly available data of the second {\em Fermi}-LAT pulsar catalog.\footnote{\url{http://fermi.gsfc.nasa.gov/ssc/data/access/lat/2yr_catalog/}} Data consist of a number of $N_{\rm bin}\sim 5-10$ energy bins (each one with its extremes $E_1$ and $E_2$, and its weighted central energy $E_{\rm cent}$), each one with its associated photon flux, $F^{\rm bin}_{\rm obs}$ (in units photons~cm$^{-2}$s$^{-1}$), and its associated statistical error, $\delta F^{\rm bin}_{\rm obs}$. We consider the isotropic luminosity, $L^{\rm bin}_{\rm obs}=4\pi d^2 F^{\rm bin}_{\rm obs}$, where $d$ is the distance of the pulsar to the Earth, and we neglect possible beaming effects which would reduce the inferred luminosity. We neglect bins having only upper limits to the flux.

To explore the space of parameters, we span a grid of 41 different values of $E_\parallel$, logarithmically equi-spaced, covering two orders of magnitude, and a grid of 50 values of $x_0/R_{\rm lc} \in [0.001,1]$ (with increasing steps), plus the case of uniform effective distribution of particle. For each pair of values $(E_\parallel,x_0)$, we evaluate the expected spectrum over a grid of hundreds of points in the 100 MeV-100 GeV range. Then, we renormalize it by the best-fitting value of $N_0$, found by means of scanning a grid, having progressively finer steps, up to $dN_0/N_0 \lesssim 10^{-3}$.  Therefore, we integrate the SC photon spectrum (related to the SC energy spectrum, Eq.~\ref{eq:sed_x}, by $dN_{\rm gap}/dE=(1/E)dP_{\rm gap}/dE$) in each bin (and normalizing by the bin width), to obtain the binned SC photon spectrum (number of photons per unit energy):

\begin{equation}
L^{\rm bin}_{\rm gap} = \frac{1}{E_2-E_1} \int_{E_1}^{E_2} \frac{1}{E}\frac{dP_{\rm gap}}{dE} {\rm d}E~.
\end{equation}
Therefore, we calculate the goodness-of-fit indicator in the standard way:

\begin{equation}\label{eq:chi2}
  \tilde{\chi}^2 = \frac{1}{N_{\rm bin}-1} \sum_{\rm bin} \frac{(L^{\rm bin}_{\rm obs} - L^{\rm bin}_{\rm gap})^2}{(\delta L^{\rm bin}_{\rm obs})^2}~,
\end{equation}
where $\delta L^{\rm bin}_{\rm obs} = 4\pi d^2 \delta F^{\rm bin}_{\rm obs}$ is the luminosity error for each bin. Such error neglects the (likely large) uncertainties on distance and beaming factor, but this would only change the luminosity, thus providing a different inferred normalization $N_0$, but without changing the spectral shape, and therefore, the constrains on $E_\parallel$ and $x_0/R_{\rm lc}$.

In the plots of the paper, following the literature, we plot the binned functions $E^2dN/dE \equiv E_{\rm cent}^2 L^{\rm bin}$, in units erg~s$^{-1}$, both for the theoretical models and data.

\end{document}